\documentstyle[preprint,aps,epsfig]{revtex}
\tightenlines
\newcommand{\nn}{\nonumber}

\newcommand{\sbra}[1] { \left( #1 \right)}
\newcommand{\mbra}[1] { \left\{ #1 \right\}}
\newcommand{\lbra}[1] { \left[ #1 \right]}

\newcommand{\beqn}{\begin{eqnarray}}
\newcommand{\eeqn}{\end{eqnarray}}

\begin{document}

\preprint{\parbox{5cm}{KANAZAWA~99-26}}

\title{An Almost Perfect Quantum Lattice Action for Low-energy
$\mathbf{SU(2)}$ Gluodynamics}

\author{
Maxim~N.~Chernodub$^{a}$,
Shouji~Fujimoto$^{b}$,
Seikou~Kato$^{b}$,
Michika Murata$^{b}$,
Mikhail~I.~Polikarpov$^{a}$ 
and Tsuneo~Suzuki$^{b}$ 
}

\address{$^{a}$ ITEP, B.Cheremushkinskaya 25, Moscow 117259, Russia}
\address{$^{b}$ Institute for Theoretical Physics, Kanazawa University,
 Kanazawa 920-1192, Japan}

\date{\today}

\maketitle

\begin{abstract}
We study various representations of infrared effective theory of
$SU(2)$ Gluodynamics as a (quantum) perfect lattice action. 
In particular we derive a monopole action and a string model 
of hadrons from $SU(2)$ Gluodynamics. These are lattice actions which 
give almost cut-off independent physical quantities even on coarse 
lattices. 
The monopole action is determined by numerical
simulations in the infrared region of $SU(2)$ Gluodynamics.
The string model of hadrons is derived from the monopole action
by using BKT transformation. We illustrate the method and evaluate
physical quantities such as the string tension and the mass of the lowest 
state of the glueball {\bf analytically} using the string model of
hadrons. It turns out that the {\it classical} results
in the string model is near to the one 
in {\it quantum} $SU(2)$ Gluodynamics.

\vspace{3mm}
%\\
$PACS$: 12.38.Gc, 11.15.Ha\\
$Keywords$: lattice $SU(2)$ Gluodynamics; abelian projection; 
abelian monopole; 
block-spin transformation; perfect action; hadronic string model 
\end{abstract}

%\narrowtext
%\newpage
%%%%%%%%%%%%%%%%%%%%%%%%%%%%%%%%%%%%%%%%%%%%%%%%%%%%%%%%%%%%%%%%%%
%      I. INTRO
%%%%%%%%%%%%%%%%%%%%%%%%%%%%%%%%%%%%%%%%%%%%%%%%%%%%%%%%%%%%%%%%%%
\section{Introduction}

Low-energy effective theory of QCD is important for analytical
understanding of hadron physics. Before derivation of such an
effective theory we have to explain
the most important non-perturbative phenomenon, quark confinement.
Wilson's lattice formulation \cite{Wilson} shows that the confinement is
a property of a non-abelian gauge theory of strong interaction. At strong
coupling the confinement is proved analytically. At weak coupling
(near to the continuum limit) there are a lot of numerical
calculations showing the confinement of color. The mechanism of confinement
is, however, still not well understood. One of approaches to the 
confinement problem
is to search for relevant dynamical variables and
to construct an effective theory in terms of these variables.

\mbox{}From this point of view the idea proposed by 't Hooft
\cite{'thooft} is very promising. It is based on the fact
that after a partial gauge fixing (abelian projection) $SU(N)$
gauge theory is reduced to an abelian $U(1)^{N-1}$ theory with $N-1$
different types of abelian monopoles. Then the
confinement of quarks can be explained as the dual Meissner effect
which is due to condensation of these monopoles.
The QCD vacuum is dual to the ordinary
superconductor: the monopoles playing the role of
the Cooper pairs. The confinement occurs due to the formation of a
string with electric flux between quark and anti-quark. It
is a dual analogue of the Abrikosov string \cite{Abrikosov}. The
mechanism of confinement is usually called the dual
superconductor mechanism.

There are many ways to perform the abelian projection, but in the
Maximal abelian (MA) gauge~\cite{Kronfeld} many numerical results
support the dual superconductor picture of confinement \cite{domi} in
the framework of lattice Gluodynamics (see, for example, reviews
\cite{Reviews,ChPoRev}). These results suggest that the abelian
monopoles which appear after the abelian projection of QCD, are
relevant dynamical degrees of freedom in the infrared (IR) region.  We
expect hence, after integrating out all degrees of freedom other than
the monopoles, an effective theory described by the monopoles works
well in the IR region of Gluodynamics.

The effective monopole action on the MA projection of $SU(2)$ lattice
Gluodynamics was obtained by Shiba and Suzuki \cite{shiba_suzuki}
using an inverse Monte-Carlo method ~\cite{Swendsen}. Assuming that
the lattice action contains only quadratic terms of monopole
currents, they found that the action has a form theoretically
predicted by Smit and van der Sijs \cite{SvdS}. This was the first
derivation of an effective theory of lattice Gluodynamics in terms of
the monopole currents.  However the steps of block-spin transformation
performed in \cite{shiba_suzuki} were rather few to see the continuum limit.  
%-------------------------------------------------------
In \cite{nakam} they considered also four- and six-point interactions
assuming a direction symmetric action on the large ($48^4$) lattice.
More steps of the block-spin transformations were carried out also.  
It is stressed that the action
seems to satisfy a scaling behavior, that is, it depends on the
physical length $b = n a(\beta)$ alone, where $n$ is the number of the
blocking transformations and $a(\beta)$ is the lattice spacing. This
remarkable scaling is consistent with the behavior of the
perfect action on the renormalized trajectory (RT) which is an effective
theory in the continuum limit formulated on the lattice with the
lattice distance $b$.  Here $b$ plays a role of the physical scale at
which the effective theory is considered. 
On RT, although we
can predict physical quantities only on the $b$ lattice sites, they
are the same as evaluated from the continuum theory. For example, the
continuum rotational invariance should be satisfied.
%--------------------------------------------------------------
The restoration of the continuum rotational invariance for the 
quark-antiquark static potential was studied using a naive 
Wilson loop operator. 
However, the continuum rotational invariance was not confirmed
in the IR region of SU(2) Gluodynamics\cite{kato98}.
This is because the cut-off effect of such an operator is of order of the 
lattice spacing of the coarse lattice. 
To check restoration of the continuum rotational invariance, 
we should determine the correct form of physical operators
(the perfect operator) as well as the perfect action on the 
blocked lattice.

%----------------------------------------------------------------
 The main task of this publication is to derive the perfect 
monopole and the string action as an low-energy effective theory of
 SU(2) Gluodynamics and evaluate physical quantities analytically
using a renormalized operator.

In Section~\ref{PerfecAc} we discuss how to derive the renormalized
monopole and the string action from SU(2) Gluodynamics. We show 
new results of the analysis of the monopole action which
is obtained by using inverse Monte-Carlo method.
In Section~\ref{PfecOpe} we discuss how to construct the 
perfect operator for the static potential.
In Section~\ref{NumRes} we calculate the string tension and
the glueball mass for the SU(2) Gluodynamics in terms of the 
strong coupling expansion of
the string model analytically. It turns out that the {\it classical} 
results in the string model is near to the one 
in {\it quantum} $SU(2)$ Gluodynamics. The continuum rotational
invariance of the static potential is shown also analytically.
In Section~\ref{AN} we analyse the numerical results
in details.
Section~\ref{concl} is devoted to concluding remarks.

%%%%%%%%%%%%%%%%%%%%%%%%%%%%%%%%%%%%%%%%%%%%%%%%%%%%%%%%%%%%%%%%
%  Perfect actions for QCD
%%%%%%%%%%%%%%%%%%%%%%%%%%%%%%%%%%%%%%%%%%%%%%%%%%%%%%%%%%%%%%%%
\section{Almost perfect monopole action from SU(2) Gluodynamics}
\label{PerfecAc}

%In this section we illustrate the method to derive 
%the monopole from SU(2) Gluodynamics.

%%%%%%%%%%%%%%%%%%%%%%%%%%%%%%%%%%%%%%%%%%%%%%%%%%%%%%%%%%%%%%%%%%
%  monopole action (numerical study)
%%%%%%%%%%%%%%%%%%%%%%%%%%%%%%%%%%%%%%%%%%%%%%%%%%%%%%%%%%%%%%%%%%
\subsection{Our method}
\label{foom}
%\input{IMM}
%-----------------------------------------
The method to derive the monopole action is the following:

\begin{itemize}
\item[1] 
We generate $SU(2)$ link fields $\mbra{U(s,\mu)}$ using the
simple Wilson action for SU(2) Gluodynamics.
We consider $24^4$ and $48^4$ hyper-cubic lattice for $\beta=2.0\sim 2.8$.
\item[2] 
Next we perform an abelian projection in the Maximal
abelian gauge to separate abelian link variables $\mbra{u(s,\mu)
= e^{i\theta_{\mu}(s)}}(-\pi\le \theta_\mu(s) <\pi)$ from
gauge fixed $SU(2)$ link fields.
\item[3] 
Monopole currents can be defined from abelian plaquette
variables $\theta_{\mu\nu} (s)$ following DeGrand and Toussaint\cite{D_T}.
The abelian plaquette variables are written by
\begin{eqnarray}
\theta_{\mu\nu}(s)&\equiv&\theta_{\mu}(s)+\theta_{\nu}(s+\hat{\mu})
                   -\theta_{\mu}(s+\hat{\nu})-\theta_{\nu}(s),
\quad (-4\pi< \theta_{\mu\nu}(s) < 4\pi).
\end{eqnarray}
It is decomposed into two terms:
\begin{eqnarray}
\theta_{\mu\nu}(s)&\equiv&\bar\theta_{\mu\nu}(s)+2\pi{n}_{\mu\nu}(s),
\qquad (-\pi \le \bar\theta_{\mu\nu}(s) < \pi).
\end{eqnarray}
Here, $\bar\theta_{\mu\nu}(s)$ is interpreted as the electro-magnetic
flux through the plaquette and
the integer $n_{\mu\nu} (s)$ corresponds to the number of Dirac string
penetrating the plaquette.
One can define quantized conserved monopole currents
\begin{eqnarray}
k_\mu(s)=\frac{1}{2}\epsilon_{\mu\nu\rho\sigma}\partial_\nu
                           n_{\rho\sigma}(s+\hat{\mu}), 
\label{eqn:km}
\end{eqnarray}
where $\partial$ denotes the forward difference on the 
lattice.
The monopole currents satisfy a conservation law
$\partial^{\prime}_{\mu}k_\mu(s) = 0$ by definition,
where $\partial^{\prime}$ denotes the backward difference on the 
lattice.
\item[4]
We consider a set of independent and local monopole interactions 
which are summed up over
the whole  lattice. We denote each operator as ${\cal S}_i[k]$.
Then
the monopole action can be written as a linear combination of these
operators:
\begin{eqnarray}
 {\cal S}[k] = \sum_{i} G_i {\cal S}_i[k],
\end{eqnarray}
where $G_i$ are coupling constants.

We determine the set of couplings $G_i$ from the monopole current 
ensemble $\mbra{k_{\mu}(s)}$ with the aid of an inverse 
Monte-Carlo method first developed by Swendsen 
and extended to closed monopole currents by Shiba and Suzuki 
~\cite{shiba_suzuki,Swendsen}.

Practically, we have to restrict the number of interaction terms. 
It is natural to assume that monopoles which are far apart do not 
interact strongly and to consider only short-ranged interactions 
of monopoles. The form of actions adopted here
is 27 quadratic interactions and 4-point and 6-point interactions.
We have not assumed a direction symmetric form of the action as done
in \cite{nakam}.
The detailed form of %quadratic 
interactions are shown in Appendix A.
Note that all possible types of interactions are not independent due to
the conservation law of the monopole current. So we get rid of
almost all the perpendicular interactions by the use of 
the conservation rule.
 The validity of the truncation has been studied and supported in 
the earlier works.
For details, see \cite{shiba_suzuki,nakam}.

\item[5] 
We perform a block-spin transformation in terms of 
the monopole currents on the dual lattice 
to investigate the renormalization flow in the IR region. 
We adopt $n=1,2,3,4,6,8$ extended conserved monopole currents
as an $n$ blocked operator\cite{ivanenko}:
\begin{eqnarray}
K_\mu(s^{(n)})&=&
  \sum_{i,j,l=0}^{n-1}
    k_\mu(ns^{(n)}+(n-1)\hat{\mu}+i\hat{\nu}+j\hat{\rho}+l\hat{\sigma})\\
&\equiv&{\cal B}_{k_\mu}(s^{(n)}),
\label{pfac:2}
%\label{eqn:kem}
\end{eqnarray}
The renormalized lattice spacing is $b=na(\beta)$ and the continuum limit is 
taken as the limit $n \to \infty$ for a fixed physical length $b$.

We determine the effective monopole action from the blocked 
monopole current ensemble $\mbra{K_{\mu}(s^{(n)})}$.
Then one can obtain the renormalization flow in the coupling constant space.

\item[5]
The physical length $b=n a(\beta)$ is taken in unit of the physical 
string tension $\sqrt{\sigma_{phys}}$. We evaluate 
the string tension $\sigma_{Lat}$ from the monopole part of the abelian
Wilson loops for each $\beta$ since the error bars are small in 
this case. 
The lattice spacing $a(\beta)$ is given
by the relation $a(\beta)=\sqrt{\sigma_{Lat}/\sigma_{phys}}$ \cite{nakam}.
Note that 
$b=1.0 \sigma_{phys}^{-1/2}$ corresponds to $0.45fm$, when we 
assume $\sigma_{phys}\cong (440 MeV)^2$.
\end{itemize}

%------------------------------------------
\subsection{Numerical results}
\label{resul}

We list new results below in comparison with earlier numerical
analysis of the monopole action.

\begin{figure}
\begin{center}
\begin{minipage}{80mm}
  \epsfig{file=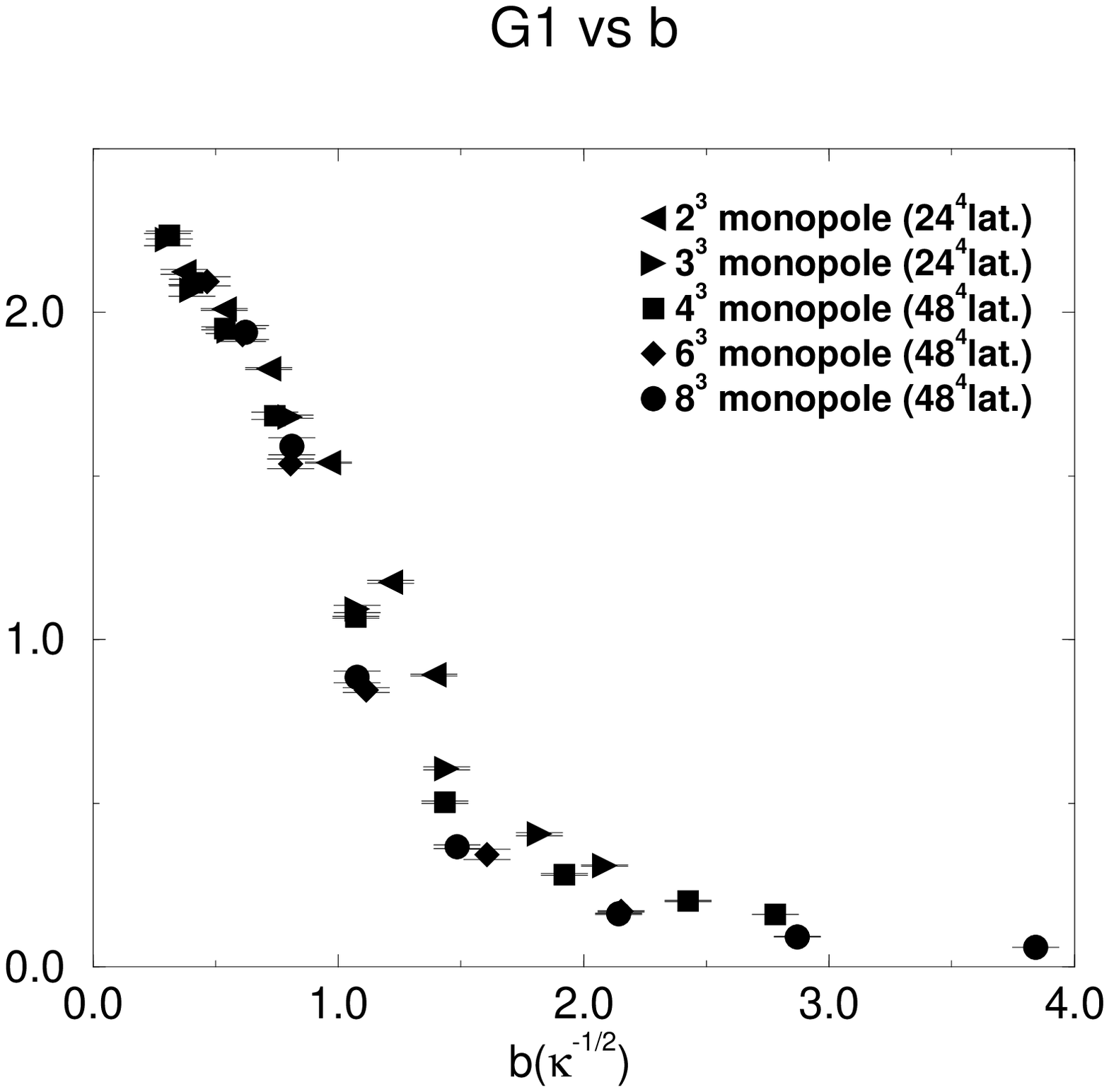,width=77mm}
\end{minipage}
\begin{minipage}{80mm}
  \epsfig{file=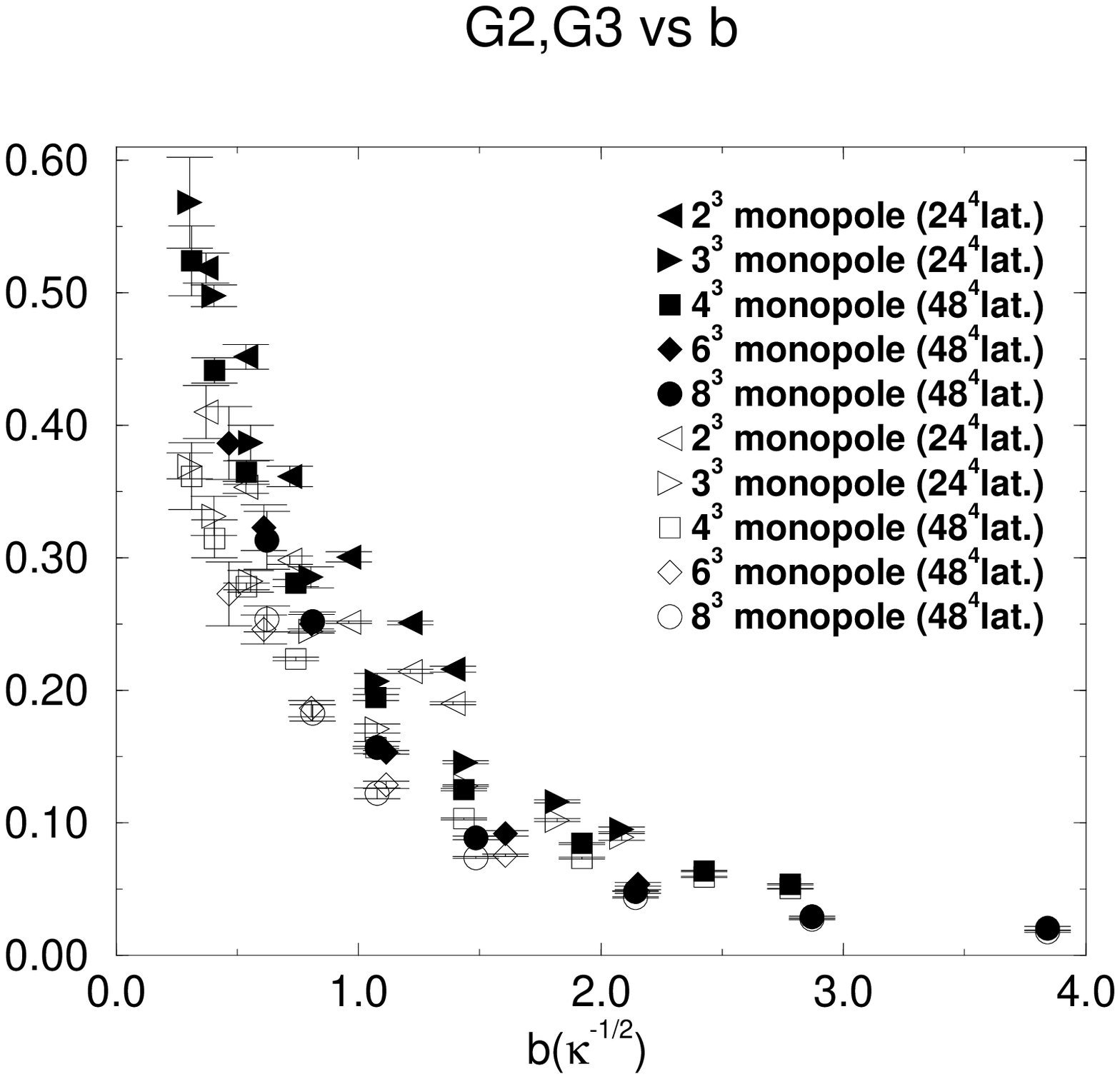,width=77mm}
\end{minipage}

\vspace{5mm}

\begin{minipage}{80mm}
  \epsfig{file=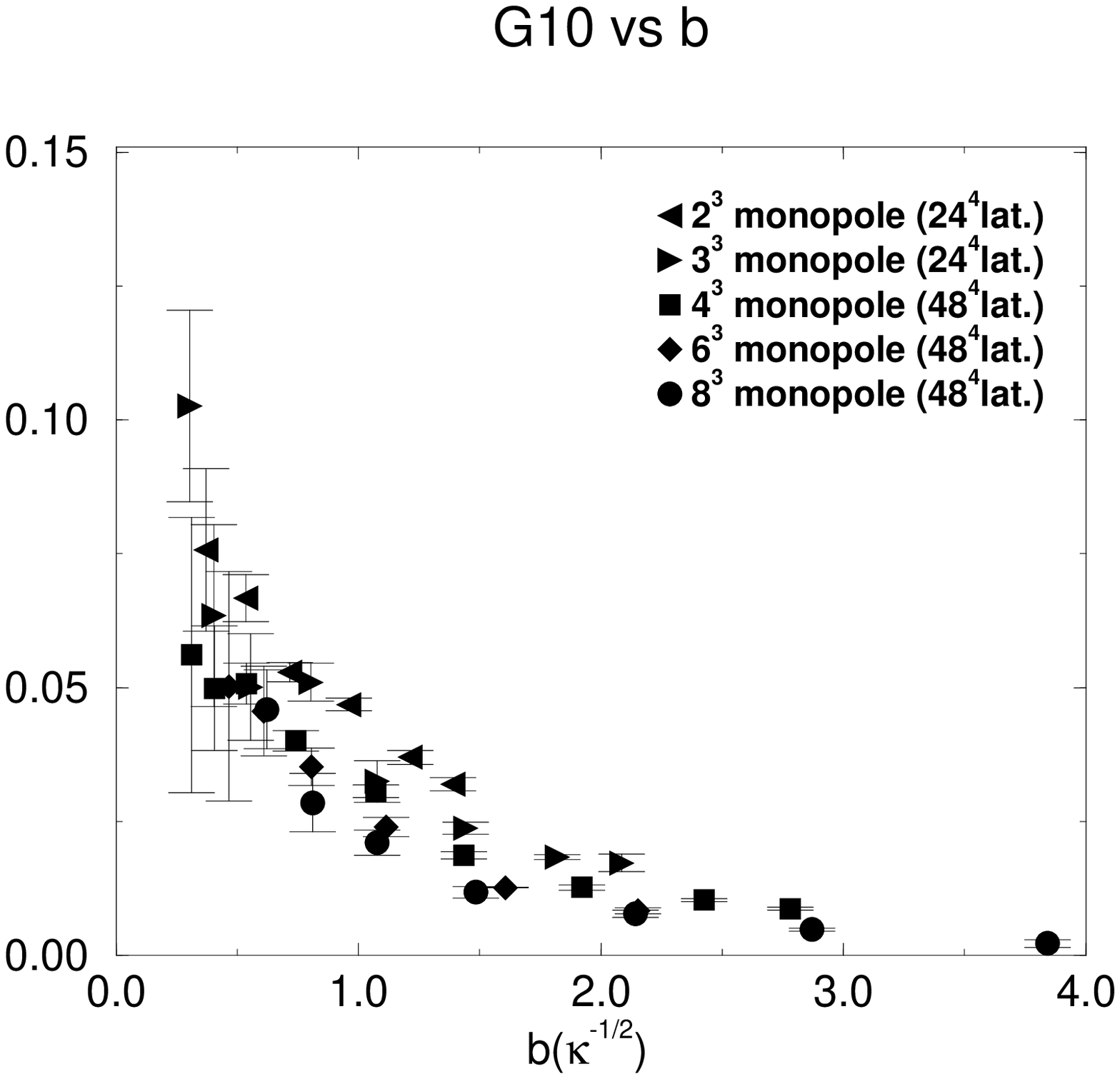,width=77mm}
\end{minipage}
\begin{minipage}{80mm}
  \epsfig{file=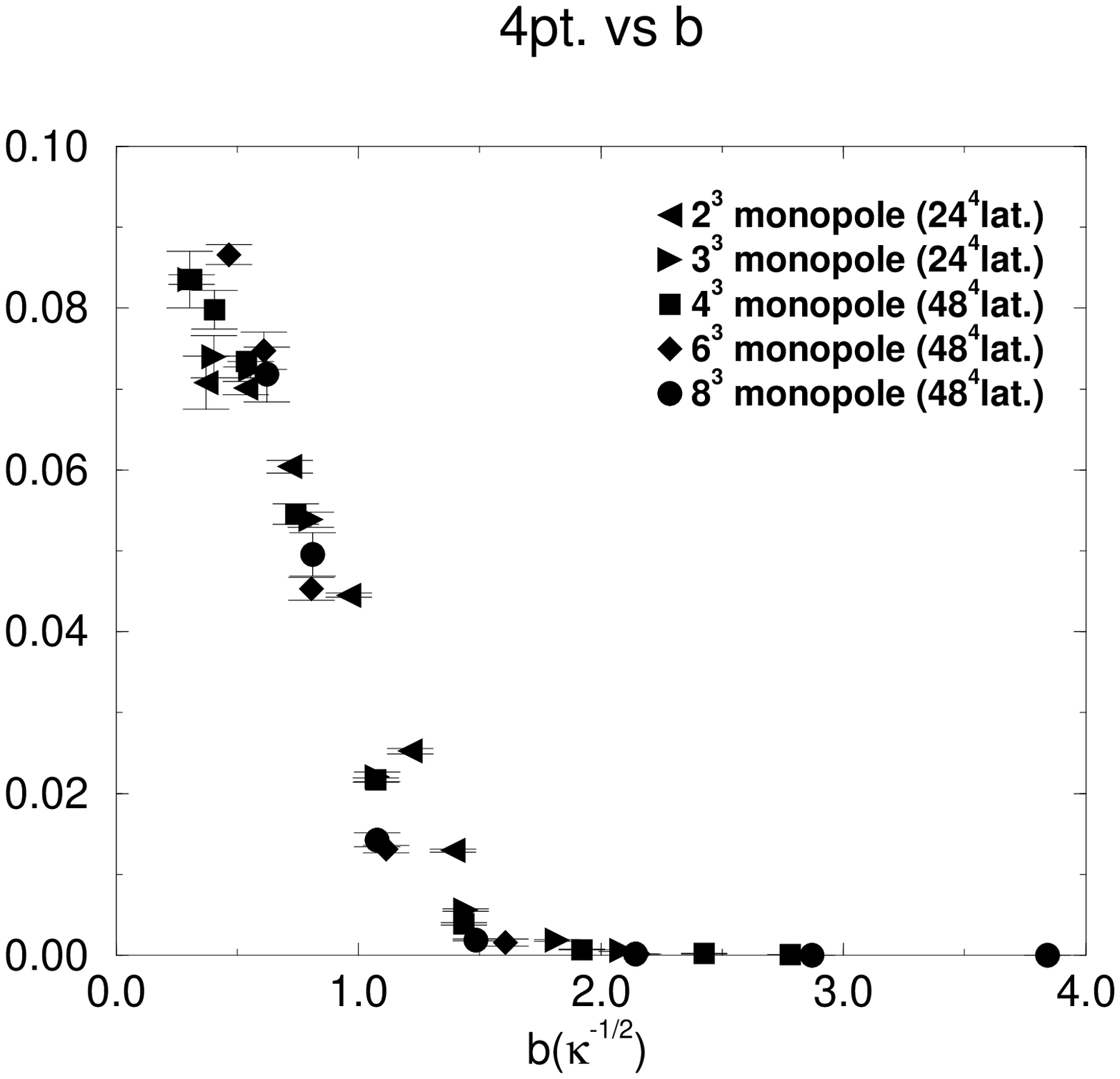,width=77mm}
\end{minipage}
\end{center}
\caption{
The couplings of quadratic interaction term and 4-point interaction 
 term versus physical length $b$.
}
\label{fig:g1-4p}
\end{figure}

\begin{figure}
\begin{center}
\begin{minipage}{80mm}
  \epsfig{file=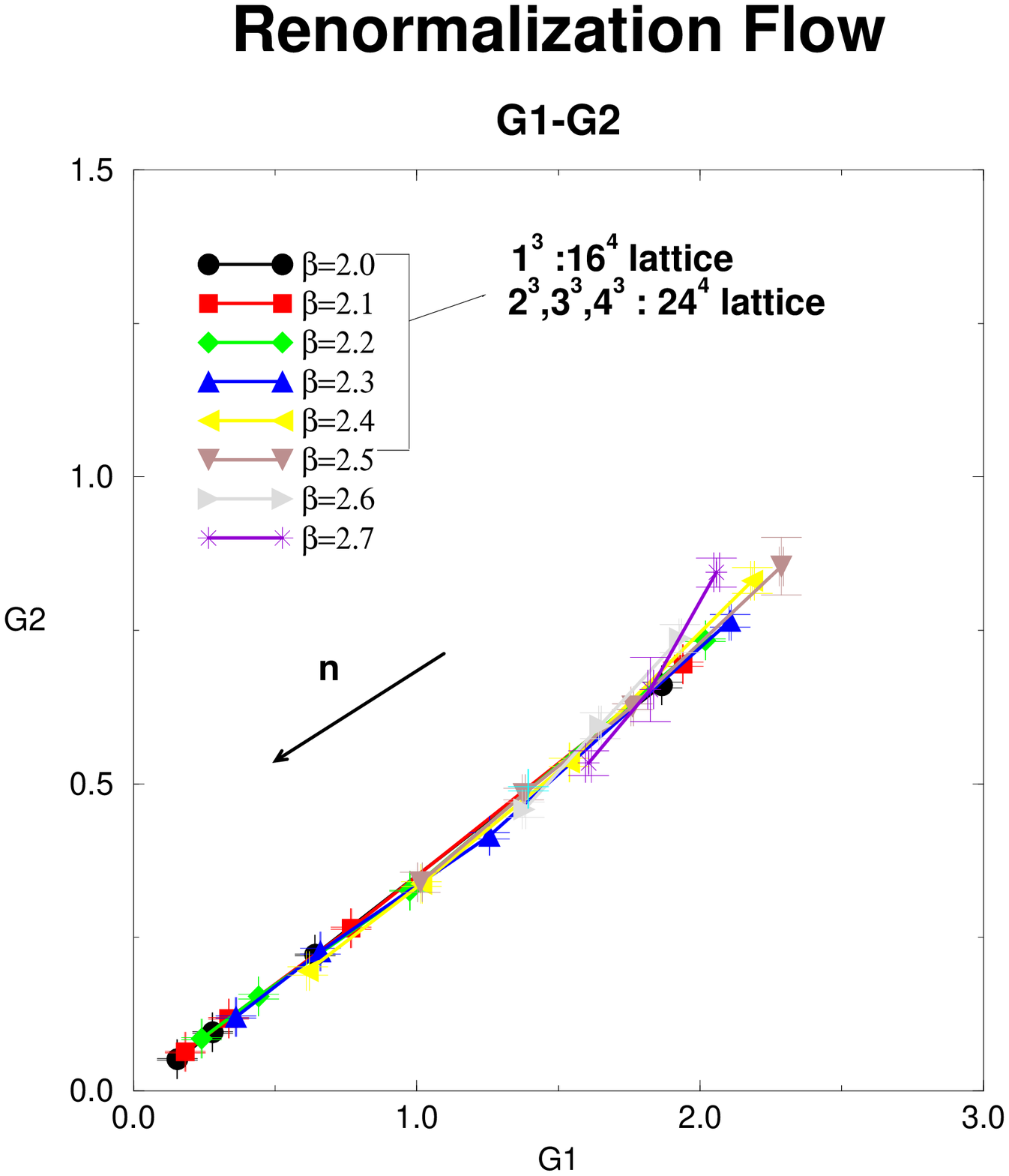,width=80mm,height=100mm}
\end{minipage}
\begin{minipage}{80mm}
  \epsfig{file=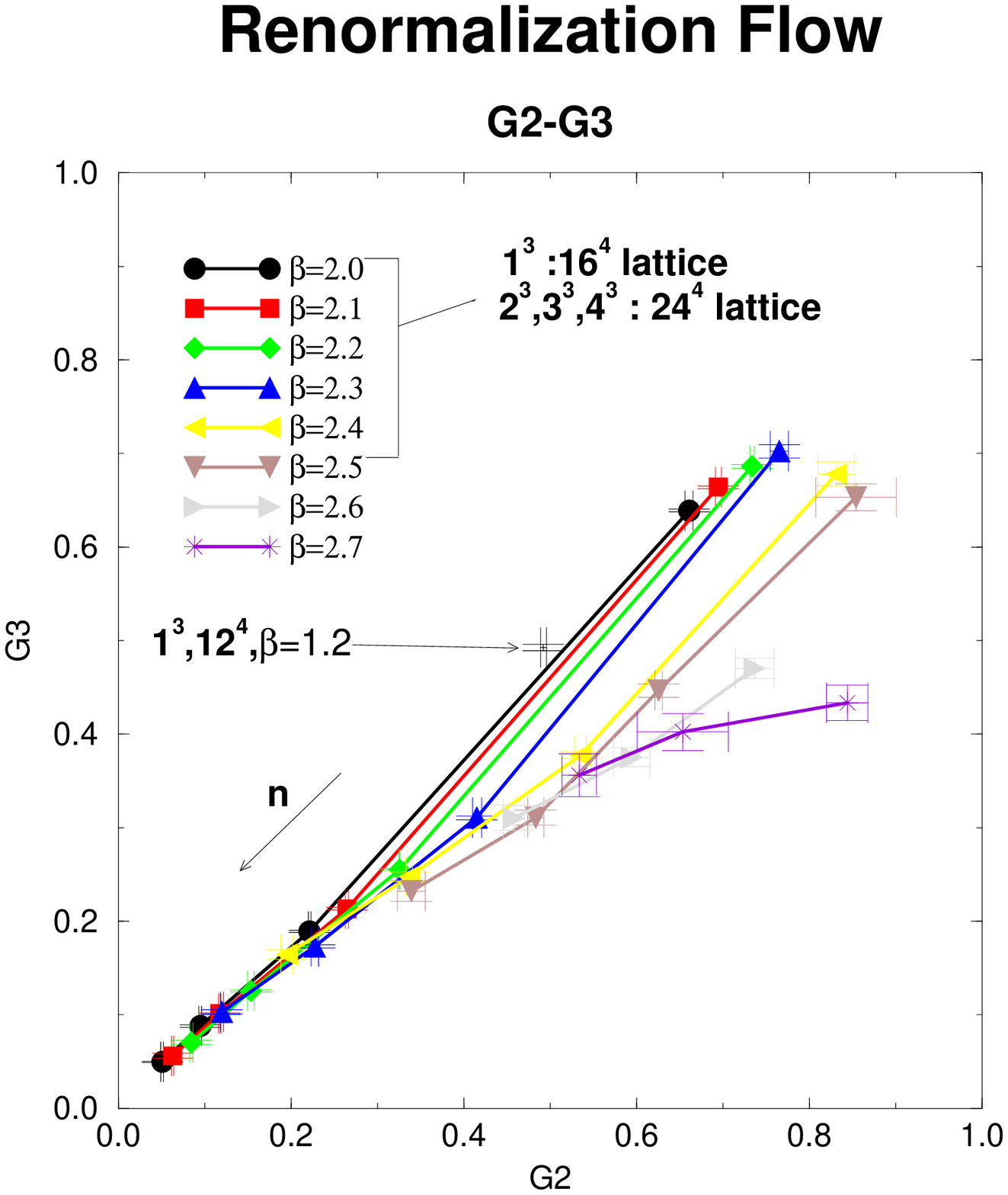,width=80mm,height=100mm}
\end{minipage}

\vspace{5mm}

\begin{minipage}{80mm}
\epsfig{file=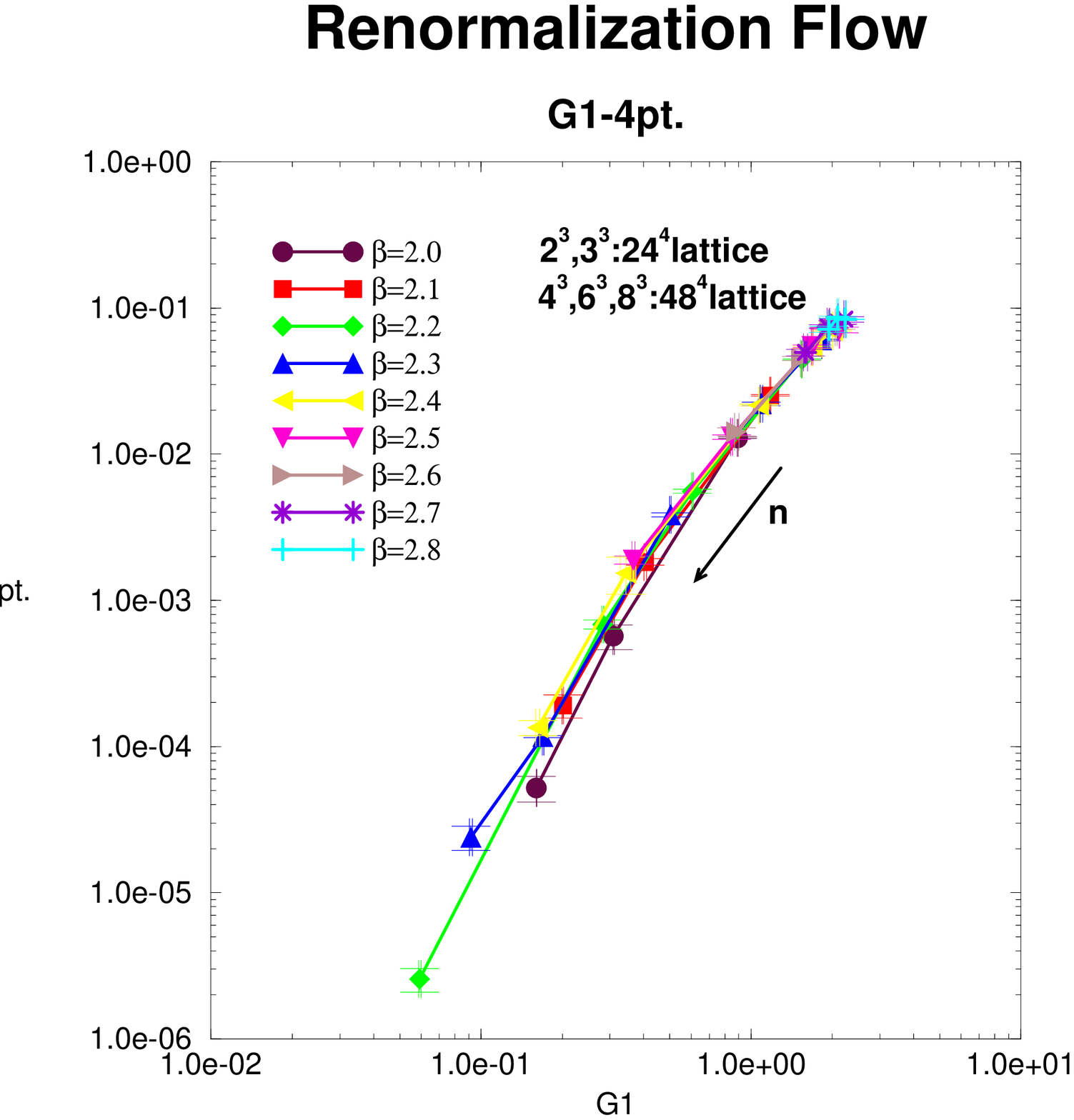,width=80mm,height=100mm}
\end{minipage}
\end{center}

\caption{
The renormalization flow on the projected plane.
}
\label{fig:RGflow}
\end{figure}

\begin{center}
\begin{figure}
\hspace{13mm}
\epsfig{file=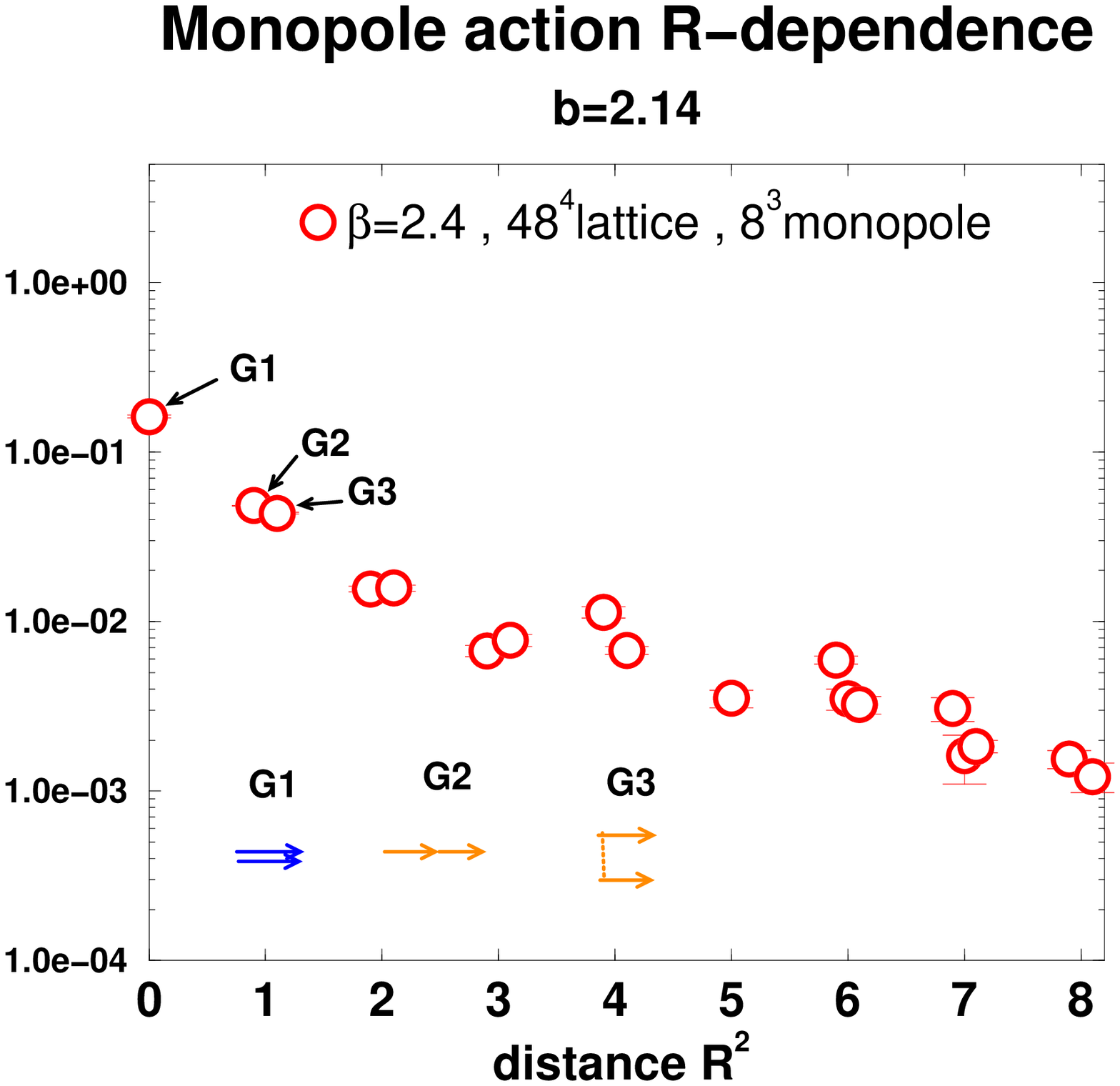,width=100mm}
\caption{
The distance dependence of the couplings of quadratic interaction 
terms at $b=2.14$.
}
\label{fig:dist}
\end{figure}
\end{center}

%--------------------------------------------
\begin{itemize}
\item[1]
The inverse Monte-Carlo method works well and the coupling constants of
the action are fixed beautifully.
The quadratic coupling constants and 4-point coupling constant
are plotted versus the physical length $b=n a(\beta)$ for each $n$ extended
monopole in Figure~\ref{fig:g1-4p}.
The first three figures show quadratic self coupling $G_1(b)$, quadratic
nearest-neighbor couplings ($G_2(b)$ (black symbol), $G_3(b)$ (open
symbol)) and $G_{10}(b)$, respectively.
The self-coupling term is dominant and the coupling constants decrease
rapidly as the distance between the two monopole currents increases.
\begin{eqnarray*}
 G_1(b) >> G_2(b) \sim G_3(b) > \cdot\cdot\cdot > G_{10}(b) > \cdot\cdot\cdot 
\end{eqnarray*}
The 4-point coupling constant becomes negligibly small in comparison 
with the quadratic couplings for large $b$ region 
($b>1.5 \sigma_{phys}^{-1/2}$). The 6-point coupling constant behaves 
similarly as the 4-point coupling does and becomes much smaller for large 
$b$ region.
\begin{eqnarray*}
 \mbox{quadratic couplings} >> \mbox{4-point coupling} 
>> \mbox{6-point coupling}
\end{eqnarray*}
>From these figures we see a scaling of the action
${\cal S}[k_{\mu}, n,a(\beta)] \to {\cal S}[K_{\mu},b=na(\beta)]$
for fixed physical length $b=na(\beta)$ looks almost good for $n \geq 4$. 
The obtained action appears to be a good approximation 
of the action on the RT.

\item[2]
In Figure~\ref{fig:RGflow} we plot the projected lines ($G_1(b)-G_2(b)$,
$G_2(b)-G_3(b)$ and $G_1(b)$-4-point, respectively) of the
renormalization flow. 
Each flow line for smaller $\beta$ ( which corresponds to 
larger $b$ ) 
is beautifully straight with very small errors.
The quadratic interactions for monopoles are dominant for larger $b$ 
, that is, only the quadratic interaction subspace seems sufficient 
in the coupling space for low-energy SU(2) Gluodynamics. 
We also see the effective monopole action tends to go to the weak 
coupling region when we go to the infrared region of SU(2) Gluodynamics.
%These results are consistent with the early analysis \cite{nakam}.

\item[3] 
The quadratic coupling constants at $b=2.14$ are plotted versus the
squared distance $R^2$ in unit of squared physical length $b^2$
in Figure~\ref{fig:dist}.
We see the direction asymmetry of the current action. 
(For example, $G_2 \ne G_3$.)
This behavior of the action does not occur in the case of compact
QED, because the monopole action can be obtained from the Villain form
of compact QED exactly in an analytical way and it does not depend
on the direction between two monopole currents.
In \cite{nakam} they have neglected this effect and have considered 
a direction symmetric form of the monopole action but as 
we will see later that this direction asymmetry of the current action
is natural and important features of the perfect lattice action.

\end{itemize}

%In this way, we determined the monopole action from SU(2) Gluodynamics. 
%Up to this stage, we used the numerical simulation. 

%However it is difficult to evaluate the static 
%potential using obtained effective monopole action by analytical
%treatment such as perturbative method. Furthermore, one can not derive
%the static potential numerically in the large $b$ region, because 
%Wilson loops become very small numerically.
%So we should consider more suitable model which is equivalent with the
%effective monopole action. This is issue of next subsection.

%%%%%%%%%%%%%%%%%%%%%%%%%%%%%%%%%%%%%%%%%%%%%%%%%%%%%%%%%%%%%%%
%   SECTION 3. A perfect operator for a physical quantities
%%%%%%%%%%%%%%%%%%%%%%%%%%%%%%%%%%%%%%%%%%%%%%%%%%%%%%%%%%%%%%%
\section{A perfect operator for physical quantities}
\label{PfecOpe}

In previous section we have studied the renormalized monopole action 
${\cal S}[k]$ performing block spin transformation up to $n=8$ 
numerically, and have found the scaling for fixed physical length $b$ 
looks almost good. 
If the continuum rotational invariance of physical observables is 
satisfied in addition in the framework of ${\cal S}[k]$,
we can regard ${\cal S}[k]$ as a good approximation of RT.

\subsection{Improved and perfect operator}
\label{IO}

In Gluodynamics, the string tension from the 
static potential is one of 
important physical quantities. However,
it is a problem 
how to evaluate the static potential between electrically charged
particles after abelian projection.
In the earlier work \cite{kato98} we considered a naive 
abelian Wilson loop operator and ${\cal S}[k]$ 
on the coarse lattice to evaluate the static potential, but 
the continuum rotational invariance
of the potential could not be well reproduced even for the 
infrared region of SU(2) Gluodynamics.
This is because the cut-off effect of such an operator is of order of the 
lattice spacing of the coarse lattice. 
Only the scaling behavior of the action is insufficient. 
We should also adopt improved physical operators  
on the coarse lattice in order to get the correct values of physical 
observables. An operator giving a cut-off independent value on RT
 is called {\bf perfect operator}.
%----------------------------------

\subsection{The method}
\label{TM}

As will be shown in Subsection~\ref{TST} ,
when we consider a monopole action composed of general quadratic 
interactions alone, a block spin transformation can be done analytically
\cite{ourpaper}. We find a perfect operator for a static potential
starting from an operator in the continuum limit.
The continuum rotational invariance is shown exactly 
with the operator. This is an example of a perfect operator.

\begin{center}
\begin{figure}
\hspace{13mm}
\epsfig{file=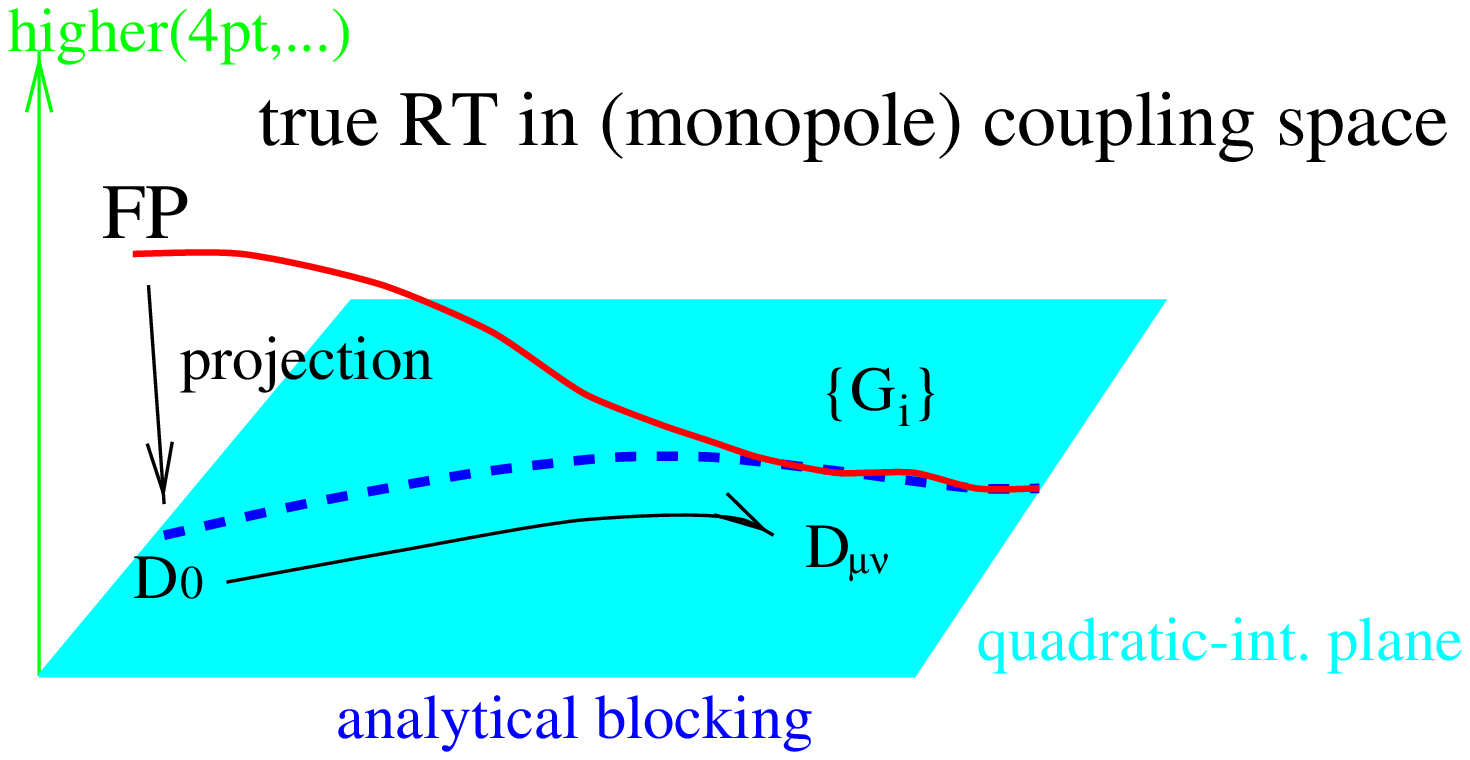,width=100mm}
\caption{
}
\label{fig:prj}
\end{figure}
\end{center}

%----------------------------------
What happens in low-energy SU(2) Gluodynamics? It is 
natural that one can not perform 
a block spin transformation analytically.
However, as shown in the previous section, the abelian monopole 
action ${\cal S}[k]$
which is obtained numerically is well approximated by 
quadratic interactions alone for large $b$. The monopole action on the 
renormalized trajectory (RT) is
expected to be  
near to the quadratic coupling constant plane in the infrared region.
We can perform the analytic block spin
transformation along the flow projected on the quadratic coupling
constant plane as shown in Figure~\ref{fig:prj}.
When we define an operator on the fine $a$ lattice, 
we can find a perfect operator along the projected flow 
in the $a\to 0$ limit for fixed $b$.
{\it Let us adopt the perfect operator on the projected space 
as an approximation of the correct
operator for the action ${\cal S}[k]$ on the coarse $b$ lattice.}
It will be shown in the following Subsection~\ref{OAC} that the
above standpoint may be justified as long as the quadratic monopole
interactions are dominant.

\subsection{Various operators for a static potential}
\label{VOSP}

%----------- cQED ---------------------
There is another problem what is 
the correct operator for the abelian static potential in 
abelian projected SU(2)
Gluodynamics on the fine a lattice.
First let us 
consider the following abelian gauge theory of the generalized Villain
form on a fine lattice with a very small lattice distance:
\beqn
{\cal S}[\theta,n] = \frac{1}{4\pi^2}\sum_{s,s';\mu>\nu}
(\partial_{[\mu}\theta_{\nu]}(s)+2\pi n_{\mu\nu}(s))
(\Delta_L D_0)(s-s')
(\partial_{[\mu}\theta_{\nu]}(s')+2\pi n_{\mu\nu}(s')),
\label{ap1:1}
\eeqn
where $\theta_{\mu}(s)$ is a compact abelian gauge field
and the integer-valued tensor $n_{\mu\nu}(s)$ comes from
the periodicity of the lattice action (\ref{ap1:1}).
Both of the variables are defined on the original lattice.
$\Delta_L(s-s')=-\partial\partial^{'}\delta_{s,s'}$ 
is the lattice Laplacian and we write
$D_0 = \beta \Delta_L^{-1} + D_0^{'}$ for later convenience,
where $D_0^{'}$ is a general operator. Since we are considering a
fine lattice near to the continuum limit, we assume the direction 
symmetry of $D_0^{'}$.
Note that $D_0=2\pi^2\beta_V\Delta_L^{-1}$ corresponds to the ordinary 
Villain action for compact QED.
In this type of model, it is natural to use an abelian Wilson loop 
$W({\cal C})=\exp i\sum_{\cal C}(\theta_{\mu}(s),J_{\mu}(s))$
for particles with fundamental abelian charge, where $J_{\mu}(s)$ is
an abelian integer-charged electric current.
The expectation value of $W({\cal C})$ is written as 
\begin{eqnarray}
\left< W(C) \right> &=&
  \left<
    \exp\left\{
      i \sum_{s,\mu} J_{\mu}(s)\theta_{\mu}(s)
    \right\}
  \right>
= Z[J]/Z[0], \\
Z[J] &\equiv&
  \int_{-\pi}^{\pi}\prod_{s;\mu}d\theta_{\mu}(s)
  \sum_{n_{\mu\nu}(s)=-\infty}^{+\infty}
  \exp\left\{
    -{\cal S}[\theta,n] + i \sum_{s,\mu}J_{\mu}(s)\theta_{\mu}(s)
  \right\}.
%\label{ap1.1}
\end{eqnarray}

%------ DGL -----------------
Next it is known that the theory with the above action (\ref{ap1:1})
is equivalent to the lattice form of the modified London limit of the dual 
abelian Higgs model\cite{suzu89} as shown in Appendix B
\beqn
&&{\cal S}[C, \phi, l] = \frac{1}{4\beta}\sum_{s;\mu>\nu}
                       (\partial_{[\mu}C_{\nu]}(s))^2 \nonumber \\
&&\qquad  +\frac{1}{4}\sum_{s,s';\mu}
                       (\partial_{\mu}\phi(s)-C_{\mu}(s)+2\pi l_{\mu}(s))
                       D_0^{'-1}(s-s')
                       (\partial_{\mu}\phi(s')-C_{\mu}(s')+2\pi l_{\mu}(s')).  
\label{ap1:?}
\eeqn
The static potential for electrically charged
particles is evaluated by a dual 't Hooft operator
\beqn 
 H({\cal C}) = \exp\mbra{-\frac{1}{4\beta}\sum_{s;\mu>\nu}
                       (\partial_{[\mu}C_{\nu]}(s)
                          -2\pi {}^*S^J_{\mu\nu}(s))^2
                         +\frac{1}{4\beta}\sum_{s;\mu>\nu}
                       (\partial_{[\mu}C_{\nu]}(s))^2},
\label{ap1:?2}
\eeqn
where ${}^*S^J_{\mu\nu}(s)$ is dual to the surface which is spanned 
inside the contour $J_{\mu}(s)$.

%---------- mono -----------------------
Thirdly, when use is made 
of the BKT transformation\cite{Bere,KT,misha1}, the action (\ref{ap1:1}) 
is equivalent to the following monopole action
\begin{eqnarray}
S[k_{\mu}(s)]=\sum_{s,s',\mu}k_\mu(s)D_0(s-s')k_\mu(s').\label{eqn.S}
\label{ap1?3}
\end{eqnarray}
We see that the area law term is given correctly also by 
the following operator in the monopole representation as 
shown in Appendix B:
\begin{eqnarray}
W_m({\cal C})&=&
\exp\bigg( 2\pi i\sum_{s,\mu}N_{\mu}(s)k_{\mu}(s) \bigg),
\label{eqn.WC}\\
N_{\mu}(s)&=&\sum_{s'}\Delta_L^{-1}(s-s')\frac{1}{2}
\epsilon_{\mu\alpha\beta\gamma}\partial_{\alpha}
S^J_{\beta\gamma}(s'+\hat{\mu}), 
\label{ap1?4}
\end{eqnarray}
where $S^J_{\beta\gamma}(s'+\hat{\mu})$ is a plaquette variable satisfying 
$\partial'_{\beta}S^J_{\beta\gamma}(s)=J_{\gamma}(s)$ and the coordinate 
displacement $\hat{\mu}$ is due to the interaction between dual
variables.

However the expectation values of the above three operators are not
completely equivalent. 
When we consider infrared effective abelian theories, it is
natural that the static potential between electric charges  
becomes Coulombic in the deconfinement phase.
The 't Hooft operator in the dual abelian Higgs model or the Wilson loop
in the generalized Villain form
reproduce this behavior. However, {\it it is stressed that all three 
operators give the same area law, since the differences give only 
Coulombic or Yukawa potentials.}
Since we are interested in the  string tension, let us  consider the 
operator (\ref{eqn.WC}) from now on.
See Appendix B for details. 

\subsection{Analytic blockspin transformation}
\label{TST}

We construct a block spin transformation (\ref{pfac:2}) 
of monopole currents. \footnote{
Note that the current $K_\mu(s^{(n)})$ on the coarser lattice with a lattice 
distance $b=na$ satisfies the current conservation 
$\partial'_{\mu}K_\mu(s^{(n)})=0$ by definition.}
Integrating out the monopole current variable on the fine lattice we
arrive at an effective action and the loop operator  
for the static potential on the coarse lattice\cite{ourpaper}.
Let us start from
\begin{eqnarray}
 \langle W_m({\cal C}) \rangle
&=&  \sum_{k_{\mu}(s)=-\infty
         \atop{\partial^{\prime}_{\mu}k_{\mu}(s)=0}}^{\infty}
  \exp
  \mbra{-\sum_{s,s',\mu} k_{\mu}(s)D_0(s-s')k_{\mu}(s')
       +2\pi i\sum_{s,\mu} N_{\mu}(s)k_{\mu}(s)} \nonumber \\
&& \times \prod_{s^{(n)},\mu}\delta
      \bigg(K_{\mu}(s^{(n)})-{\cal B}_{k_\mu}(s^{(n)})\bigg)
      /{\cal Z}[k].
\label{pfac:3}
\end{eqnarray}
The cutoff effect of the operator (\ref{pfac:3}) is $O(a)$ by definition.
This $\delta$-function renormalization group transformation can be
done analytically. Taking the continuum limit $a\to 0$, $n\to \infty$
(with $b=na$ is fixed) finally, we obtain the expectation
value of the operator on the coarse lattice with spacing $b=n a$
\cite{ourpaper}: 
\begin{eqnarray}
\langle W_m({\cal C}) \rangle
&=&
  \exp\Biggl\{
    - \pi^2 \int_{-\infty}^{\infty}\!\! d^4xd^4y
    \sum_{\mu}N_{\mu}(x)D_0^{-1}(x-y)N_{\mu}(y)
\nonumber \\
&&
    + \pi^2 b^8\!\!\!\! \sum_{s^{(n)},s^{(n)'}\atop{\mu,\nu}}
    \!\!\!\!
    B_{\mu}(bs^{(n)})
      D_{\mu\nu}(bs^{(n)}-bs^{(n)'})
    B_{\nu}(bs^{(n)'})
  \Biggr\}
\nonumber \\
&&
\times \!\!\!\!
\sum_{b^3K_\mu(bs)=-\infty\atop\partial'_\mu K_\mu=0}^{\infty}
\!\!\!\!\!\!
  \exp\Biggl\{
    - S[K_{\mu}(s^{(n)})]
\nonumber \\
&&
    +2 \pi i b^8\!\!\!\! \sum_{s^{(n)},s^{(n)'}\atop{\mu,\nu}}
    \!\!\!\!
     B_{\mu}(bs^{(n)})
       D_{\mu\nu}(bs^{(n)}-bs^{(n)'})
     K_{\nu}(bs^{(n)'})
  \Biggr\}
  \Bigg/
%\nonumber \\
  \!\!\!\!
  \sum_{b^3K_\mu(bs)=-\infty\atop\partial'_\mu K_\mu=0}^{\infty}
  \!\!\!\!\!\! Z[K,0],
\label{opwil:1}
\end{eqnarray}
where
\begin{eqnarray}
B_\mu(bs^{(n)}) &\equiv&
\lim_{a\to 0 \atop{n\to\infty}}
  a^8\sum_{s,s',\nu}
    \Pi_{{\neg}\mu}(bs^{(n)}-as)
\left\{
  \delta_{\mu\nu}
  -\frac{\partial_{\mu}\partial'_{\nu}}{\sum_{\rho}\partial_{\rho}
  \partial'_{\rho}}
\right\}D_0^{-1}(as-as')
N_{\nu}(as'),
\label{opwil:9}
\\
\Pi_{\neg\mu}(bs^{n}-as)&\equiv&
\frac{1}{n^3}
  \delta\left( nas_\mu^{(n)}+(n-1)a-as_\mu \right)
%\nonumber \\
%&&\qquad\qquad\quad
  \times
  \prod_{i(\ne \mu)}\left(
    \sum_{I=0}^{n-1}\delta\left( nas_i^{(n)}+Ia-as_i \right)
  \right).
\end{eqnarray}
$S[K_{\mu}(s^{(n)})]$ denotes the effective action defined on the coarse
lattice:
\begin{eqnarray}
  S[K_{\mu}(s^{(n)})] =
  b^8 \sum_{s^{(n)},s^{(n)'}}\sum_{\mu,\nu}K_{\mu}(bs^{(n)})
  D_{\mu\nu}(bs^{(n)}-bs^{(n)'})K_{\nu}(bs^{(n)'}).
\label{pfac:5}
\end{eqnarray}
Since we take the continuum limit analytically, the operator (\ref{opwil:1})
does not have no cutoff effect.

The momentum representation of $D_{\mu\nu}(bs^{(n)}-bs^{(n)'})$
takes the form
\begin{eqnarray}
D_{\mu\nu}(p)=
  A_{\mu\nu}^{GF-1}(p) 
  - \frac{1}{\lambda}
      \frac{\hat{p_\mu}\hat{p_\nu}}{(\hat{p}^2)^2}e^{i(p_\mu-p_\nu)/2},
%\nonumber
\label{fit:2}
\end{eqnarray}
where $A_{\mu\nu}^{\prime GF^{-1}}(p)$ is the gauge-fixed 
inverse of the following operator
\begin{eqnarray}
A'_{\mu\nu}(p)\!&\equiv&\!
\left(\prod_{i=1}^4\sum_{l_i=-\infty}^{\infty} \right)
\!\!
\Biggl\{
  \!D_0^{-1}(p+2\pi l)
  \Biggl[
    \delta_{\mu\nu}\!-\frac{(p+2\pi l)_\mu(p+2\pi l)_\nu}{\sum_i(p
+2\pi l)_i^2}
  \Biggr]
  \frac{(p+2\pi l)_\mu(p+2\pi l)_\nu}{\prod_i(p+2\pi l)_i^2}
\Biggr\}\nonumber \\
&&\hspace{2cm}\times  \frac{\left(\prod_{i=1}^4\hat{p}_i \right)^2}
{\hat{p}_\mu\hat{p}_\nu}.
\label{fit:11}
\end{eqnarray}
The explicit form of $D_{\mu\nu}(p)$ is written in Ref.~\cite{ourpaper}.
%------ string -----------------------------
Performing the BKT transformation explained in Appendix B on the coarse
lattice, we can get the loop operator for the static potential 
in the framework of the string model:
\begin{eqnarray}
\langle W_m({\cal C}) \rangle
&=& \langle W_m({\cal C}) \rangle_{cl} 
%\nonumber\\
%&& 
\times \frac{1}{Z}
  \!\!\!\!
  \sum_{\sigma_{\mu\nu}(s)=-\infty
        \atop{\partial_{[\alpha}\sigma_{\mu\nu]}(s)=0}}^{\infty}
  \!\!\!\!
  \exp
  \Bigg\{
    -\pi^2\sum_{ s,s'\atop{ \mu\neq\alpha\atop{ \nu\neq\beta } } }
    \sigma_{\mu\alpha}(s)\partial_{\alpha}\partial_{\beta}'
    D_{\mu\nu}^{-1}(s-s_1)\Delta_L^{-2}(s_1-s')\sigma_{\nu\beta}(s')
\nonumber \\
&&\qquad\qquad\qquad
    -2\pi^2\sum_{s,s'\atop{\mu,\nu}}\sigma_{\mu\nu}(s)\partial_{\mu}
    \Delta_L^{-1}(s-s')B_{\nu}(s') 
  \Bigg\}.
\label{opwil:4}
\end{eqnarray}
 $\langle W_m({\cal C}) \rangle_{cl}$ is defined by
\begin{eqnarray}
\langle W_m({\cal C}) \rangle_{cl}
&=&
  \exp
  \Bigg\{
    -\pi^2 \int_{-\infty}^{\infty}\!\!\!\!d^4xd^4y
    \sum_{\mu}N_{\mu}(x)D_0^{-1}(x-y)N_{\mu}(y)
  \Bigg\}.
\label{opwil:5}
\end{eqnarray}

\subsection{The on-axis case}
\label{OAC}

 In the above calculation, we have introduced the source term
corresponding to the loop operator for the static potential 
on the fine $a$ lattice and have 
constructed the operator on the coarse $b$ lattice by making the
blockspin transformation. To check the validity of our analysis,
it is to be emphasized that {\it
the same string tension for the flat on-axis Wilson loop can be
obtained for $I,T \to \infty$ when we consider a naive Wilson loop
operator on the coarse $b$ lattice instead of that on the fine lattice
(\ref{eqn.WC})}.
When we consider only quadratic interactions for the monopole
action, we get the classical string tension from the large 
flat Wilson loop as follows\cite{ourpaper}:
\begin{eqnarray}
\sigma_L &=& \int_{-\pi}^{\pi}\frac{d^2p}{(2\pi)^2}
  \Delta_{\rm L}^{-2}(k_1,k_2,0,0)
  \left[
    \sin^2\frac{k_2}{2}D^{-1}(k_1,k_2,0,0;\hat{1}) 
%\right.
%\nonumber \\ 
%&& \quad\quad\quad \left. 
  +\sin^2\frac{k_1}{2}D^{-1}(k_1,k_2,0,0;\hat{2})
  \right],\!
\label{on-axis}
\end{eqnarray}
where $D$ denotes the coupling of the monopole action determined
numerically on the coarse $b$ lattice.
For $I\to\infty$ and $T\to\infty$, we can easily show that $\sigma_L$
agrees exactly with the string tension derived later from
(\ref{opwil:5}) \cite{ourpaper}. Therefore, 
our analysis is natural as long as the quadratic monopole action
is a good approximation in the IR region of SU(2) Gluodynamics.
Note that we can show both quantum fluctuation parts also coincide.

%%%%%%%%%%%%%%%%%%%%%%%%%%%%%%%%%%%%%%%%%%%%%%%%%%%%%%%%
%     ANALYTICAL RESULTS OF SU(2) GLUODYNAMICS
%%%%%%%%%%%%%%%%%%%%%%%%%%%%%%%%%%%%%%%%%%%%%%%%%%%%%%%%%
\section{Analytical results of SU(2) Gluodynamics}
\label{NumRes}

\subsection{parameter fitting}
\label{ParaFit}

As shown already, the (numerically obtained) effective monopole 
action for SU(2) Gluodynamics in the IR region is well dominated by quadratic 
interactions. Hence we regard the renormalization flow obtained in 
Subsection~\ref{TST} as 
a projection of RT to the quadratic-interaction plane as written in 
Figure~\ref{fig:prj}. We adopt the perfect operator discussed in the 
previous section as the correct one on the coarse $b$ lattice in  
the low-energy SU(2) Gluodynamics. In order to know the explicit form of the 
operator, we need first to fix $D_0(s-s')$. This can be done by comparing 
$D_{\mu\nu}(bs^{(n)}-bs^{(n)'})$ with the set of numerically
obtained coupling constants of the monopole action $\mbra{G_i(b)}$
in Sec.\ref{PerfecAc}.

We assume  
$D_0(s-s')$ in the monopole action (\ref{ap1?3}) to take 
$\bar{\alpha}\delta_{s,s'}+\bar{\beta}\Delta_L^{-1}(s-s')+
\bar{\gamma}\Delta_L(s-s')$,
where  $\bar{\alpha}$, $\bar{\beta}$ and $\bar{\gamma}$ are free parameters.
We can consider more general quadratic interactions, but as we see later,
this choice is sufficient to derive the IR region of SU(2) Gluodynamics.

The inverse operator of $D_0(p)=\bar{\alpha}+\bar{\beta}/p^2+
\bar{\gamma} p^2$ 
takes the form
\begin{eqnarray}
D_0^{-1}(p)=
\kappa
\left(
  \frac{m_1^2}{p^2+m_1^2} - \frac{m_2^2}{p^2+m_2^2}
\right),
\label{fit:1}
\end{eqnarray}
where the new parameters $\kappa$, $m_1$ and $m_2$ satisfy
$\kappa (m_1^2-m_2^2)=\bar{\gamma}^{-1}, 
m_1^2+m_2^2=\bar{\alpha}/\bar{\gamma}, 
m_1^2m_2^2=\bar{\beta}/\bar{\gamma}$.

\begin{center}
\begin{figure}
\hspace{13mm}
\epsfig{file=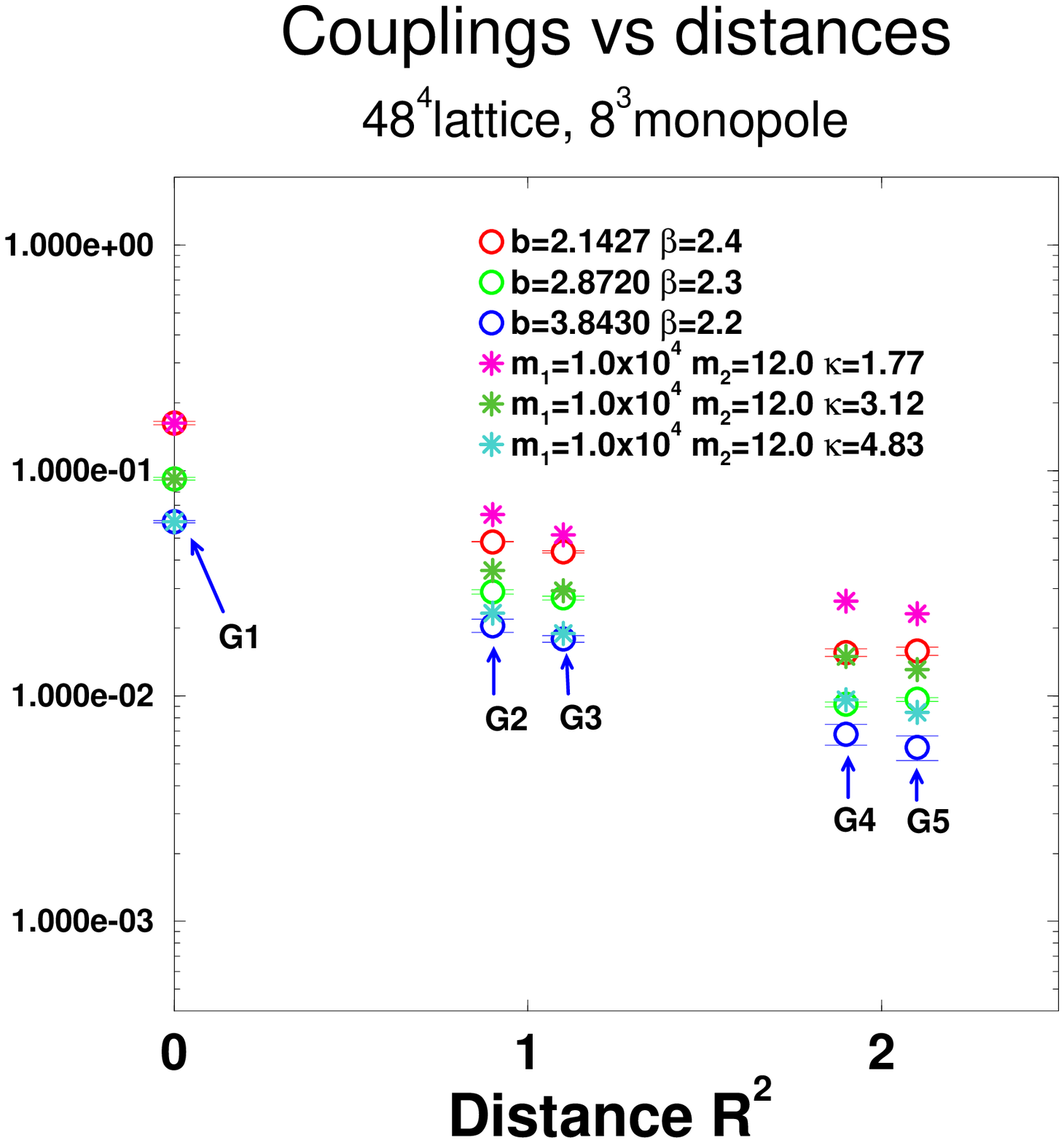,width=100mm}
\caption{
The coupling constants with the optimal values $\kappa$, $m_1$ and $m_2$ 
for $b=$ 2.1, 2.9 and 3.8 from the comparison with
numerical data.
}
\label{fig:fit}
\end{figure}
\end{center}

Substituting Eq.(\ref{fit:1}) into Eq.(\ref{fit:11}) and
performing a First Fourier transform(FFT) on the $16^4$ lattice for the 
several input values $\kappa$, $m_1$ and $m_2$ we calculate $D_{\mu\nu}(p)$.
Then one can obtain distance dependence of 
the $D_{\mu\nu}(bs^{(n)}-bs^{(n)'})$.
By matching the distance dependence of the $D_{\mu\nu}(bs^{(n)}-bs^{(n)'})$
with numerical ones, one can fit the free parameters $\kappa$, $m_1$ and $m_2$.
We find that the ratio $m_1/m_2$ is around $10^4$, but 
$m_1$ and $m_2$ can not be fixed well separately. 
Their optimal values for $b=
2.1, 2.9$ and $3.8$ %(in unit of $\sqrt{\kappa_{phys}}$) 
are given in Table~\ref{tbl:fit}, where we fix  $m_1=1.0\times 10^{4}$
and $m_2=12$ for all $b$. The coupling constants with the optimal values
are illustrated in Figure~\ref{fig:fit}.
Note that, in this figure, the lattice monopole action 
obtained from the continuum by analytical blocking also show the 
direction asymmetry. 

%--------------string tension--------------------
\subsection{The string tension}
\begin{center}
\begin{figure}
\hspace{13mm}
\epsfig{file=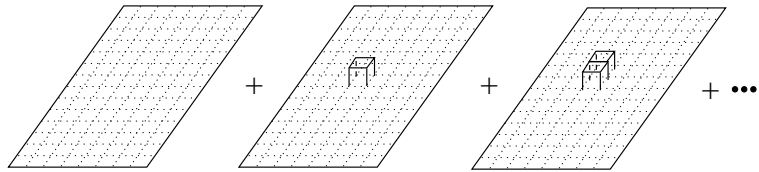,width=100mm}
\caption{The strong coupling expansion of the Wilson loop calculation.
}
\label{fig:strong}
\end{figure}
\end{center}

Let us evaluate the string tension using the perfect operator 
(\ref{opwil:4}).
The plaquette variable $S^J_{\alpha\beta}$ in Eq.(\ref{ap1?4}) for the 
static potential $V(Ib,0,0)$ is expressed by
\begin{eqnarray}
S^J_{\alpha\beta}(z)
&=&
  \delta_{\alpha 1}\delta_{\beta 4}\delta(z_{2})\delta(z_{3})
  \theta(z_{1})\theta(Ib-z_{1})
  \theta(z_{4})\theta(Tb-z_{4}).
\end{eqnarray}
In the Subsection~\ref{resul} we have seen that the monopole action on 
the dual lattice 
is in the weak coupling region for large $b$.
Then the string
model on the original lattice is in the strong coupling region. 
Therefore, we evaluate Eq.(\ref{opwil:4}) by the 
strong coupling expansion.
%------------------------------------------
The method can be shown diagrammatically in Figure~\ref{fig:strong}.
%-------------------------------------------------
\subsubsection{The classical part}
\label{clp}

As explicitly evaluated in Ref.\cite{ourpaper}, 
the classical part of the string tension coming 
from Eq.~(\ref{opwil:5}) is
\begin{eqnarray}
\sigma_{cl}=\frac{\pi\kappa}{2} \ln\frac{m_1}{m_2}.
\label{sigma_cl}
\end{eqnarray} 
$\sqrt{\sigma_{cl}/\sigma_{phys}}$ using the optimal values $\kappa$, 
$m_1$ and $m_2$ are given in Table~\ref{tbl:str}, where $\sigma_{phys}$ 
is the physical string tension.
The scaling of $\sqrt{\sigma_{cl}/\sigma_{phys}}$
for physical length $b$ seems good, although its absolute 
value is larger than 1. The difference will be analysed
later in Section~\ref{AN}.

\subsubsection{Quantum fluctuations}
\label{qf}

The next to leading quantum fluctuation term comes 
from the second part of Eq.(\ref{opwil:4}).
It corresponds to the second figure in Figure.~\ref{fig:strong} and
becomes~\cite{ourpaper}
\begin{eqnarray}
\sigma_{qf} = -\frac{4}{b^2}e^{-4\Pi(0)b^2},
\end{eqnarray}
where $\Pi(0)$ is the self coupling constant of the string action
(\ref{opwil:4}).
The total string tension is the sum
$\sigma_{tot} = \sigma_{cl} + \sigma_{qf}$.

The quantum corrections for the string tension are given in 
Table~\ref{tbl:corr}. We see they are negligibly
small in IR region of SU(2) Gluodynamics.
We can evaluate physical 
quantities
using the classical part alone in the strong coupling expansion of the 
string model.
Therefore, the strong coupling expansion works good and it is found that 
the {\it classical} string tension in the string model is near to
the one in {\it quantum} $SU(2)$ Gluodynamics.

%------------ validity
\subsubsection{The on-axis case}
\label{validity}

We evaluate next the string tension using Eq.(\ref{on-axis}),
where $D^{-1}(k)$ are determined from 
the numerical data of coupling constants. 
By using a First Fourier transform(FFT) on the $32^2$ lattice,
we perform the integration with respect to the momentum
in (\ref{on-axis}).
The results are given in 
Table~\ref{tbl:str_L}. We find that these are almost the same
as those in Table~\ref{tbl:str}.
The validity of our analysis 
in Section~\ref{PfecOpe} is confirmed.

\subsubsection{On the continuum rotational invariance}
\label{ocri}

We here comment on the continuum rotational invariance of the 
quark-antiquark static potential.
For the sake of convenience we place a pair of static quark and anti-quark
at the point $(0,0,0)$ and $(x_1, x_2, 0)$ on a three dimensional
timeslice, respectively. Both of the coordinates $x_1$ and $x_2$ denote
the sites sitting on the $b=na$ lattice. Therefore the potential becomes 
dependent only on two coordinates, $V = V(x_1, x_2)$.
In the framework of our analysis \cite{ourpaper}, 
the static potentials $V(Ib,0)$ and $V(Ib,Ib)$ can be written as
\begin{eqnarray}
V(Ib,0) &=& \frac{\pi\kappa Ib}{2} \ln\frac{m_1}{m_2},\\
V(Ib,Ib) &=& \frac{\sqrt{2}\pi\kappa Ib}{2} \ln\frac{m_1}{m_2}. 
\end{eqnarray}
The potentials from the classical part take only the linear form and the 
rotational invariance is recovered completely even for the nearest $I=1$
sites.
The recovery of the continuum rotational invariance 
of the static potential is naturally expected also for the quantum
fluctuation, since  we have introduced the source term corresponding 
to the Wilson loop on the fine $a$ lattice and we have taken the continuum
limit $a\to 0$.

%------------ glueball
\subsection{The glueball mass}

The mass spectrum in SU(2) Gluodynamics can be obtained by computing
the correlation functions of gauge invariant local operators or
Wilson loops, and looking for the particle poles.
For examples, one can consider a two point function 
of an operator ${\cal O}(t)= \sum_{\vec{x}}Tr(F^2)(\vec{x},t)$. 
For large time $t$ it is
expanded as
\begin{eqnarray}
 \langle{\cal O}(t){\cal O}(0)\rangle \simeq \sum_i c_i \exp(-M_i t),
\label{eqn:glue1}
\end{eqnarray} 
where $M_i$ is a glueball mass.

We consider here the following U(1) singlet and Weyl invariant
operator
\begin{eqnarray}
  \Psi (t) = L^{-3/2}\sum_{\vec{x}} 
                   Re \sbra{\Psi_{12} + \Psi_{23} 
                       + \Psi_{31}}(\vec{x},t)
\end{eqnarray} 
on the $a$-lattice at timeslice $t$.
Here $\Psi_{ij}(\vec{x},t)$ is an $na \times na$ abelian Wilson loop
 and $L$ stands for the linear size of the lattice.
One can check easily that this operator carries $0^{++}$ quantum number
\cite{Montovay}.
The connected two point correlation function of $\Psi$
is given by
\begin{eqnarray}
\langle\Psi (t)\cdot \Psi(0)\rangle_c 
&=&\langle\Psi(t)\cdot \Psi(0)\rangle - \langle\Psi(t)\rangle
   \langle \Psi(0)\rangle \nonumber \\
&=& \frac{6}{4V}\sum_{\vec{x},\vec{y}} \left[ 
    \left\{ \langle\Psi_{12}(\vec{x},t)\cdot \Psi_{12}(\vec{y},0)\rangle 
     + \langle\Psi_{12}(\vec{x},t)\cdot \Psi_{12}^{*}(\vec{y},0)\rangle 
    \right. \right.  \nonumber\\
&&  \left. \left.   
    - 2\langle\Psi_{12}(\vec{x},t)\rangle^2 \right\} +
2\left\{ \langle\Psi_{31}(\vec{x},t)\cdot \Psi_{12}(\vec{y},0)\rangle 
     + \langle\Psi_{31}(\vec{x},t)\cdot \Psi_{12}^{*}(\vec{y},0)\rangle 
    \right. \right.  \nonumber\\
&&  \left. \left.   
    - 2\langle\Psi_{31}(\vec{x},t)\rangle\cdot
      \langle\Psi_{12}(\vec{y},0)\rangle \right\}\right]
\label{eqn:glue3}
\end{eqnarray} 
%where we neglect the terms of the form 
%$\langle\Psi_{ij}\cdot \Psi_{kl}\rangle$
%($i=j$, $k\ne l$).
Then we evaluate each expectation value in (\ref{eqn:glue3}) by 
using the string model just as done in the case of the calculations 
of the string tension.
It turns out that the quantum correction is negligibly small 
and the classical part of the expectation value 
of the operator ${\cal O}_i$ 
(${\cal O}_1 = \Psi_{12}(\vec{x},t)\cdot \Psi_{12}(\vec{y},0)$,
${\cal O}_2 = \Psi_{12}(\vec{x},t)\cdot \Psi_{12}^{*}(\vec{y},0)$,
${\cal O}_3 = \Psi_{12}(\vec{x},t)$,
${\cal O}_4 = \Psi_{31}(\vec{x},t)\cdot \Psi_{12}(\vec{y},0)$ and
${\cal O}_5 = \Psi_{31}(\vec{x},t)\cdot \Psi_{12}^{*}(\vec{y},0)$
) in the string representation becomes
\begin{eqnarray}
\langle {\cal O}_i \rangle_{m}^{cl}
&=&
\exp
  \Bigg\{
    -\pi^2 \int_{-\infty}^{\infty}\!\!\!\!d^4xd^4y
    \sum_{\mu}N_{\mu}(x)D_0^{-1}(x-y)N_{\mu}(y)
  \Bigg\} %\\
\label{eqn:glue4}
%N_{\mu}(x)&=&\lim_{a\to 0} \sum_{y}\Delta_L^{-1}(x-y)\frac{1}{2}
%\epsilon_{\mu\alpha\beta\gamma}\partial_{\alpha}
%S_{\beta\gamma}(y+\hat{\mu})
%\label{eqn:glue5}
\end{eqnarray} 
corresponding to Eq.(\ref{opwil:5}).

\begin{center}
\begin{figure}
\hspace{13mm}
\epsfig{file=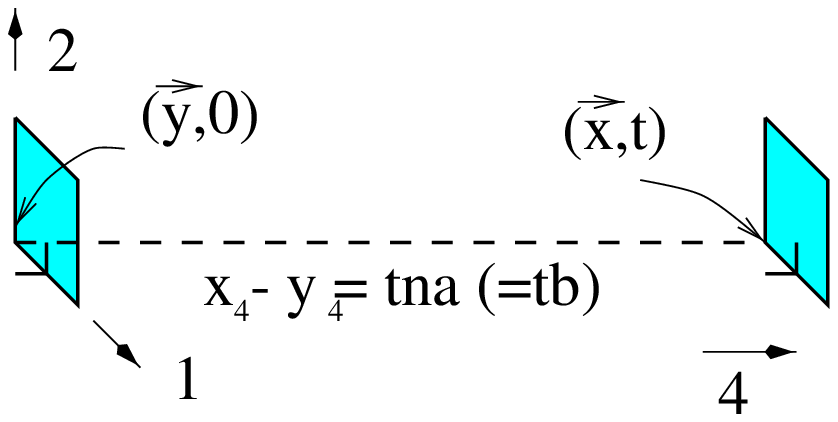,width=70mm}
\caption{
The plaquette variable $S_{\alpha\beta}$ for 
$\langle {\cal O}_1 \rangle_{m}^{cl}$.
}
\label{fig:glue}
\end{figure}
\end{center}

The plaquette variable $S_{\alpha\beta}$ in Eq.(\ref{eqn:glue4})
for $\langle {\cal O}_1 \rangle_{m}^{cl}$ is expressed by
\begin{eqnarray}
{\cal S}_{\alpha\beta}(z)&=& {\cal S}^{(1)}_{\alpha\beta}(z)
                            +{\cal S}^{(2)}_{\alpha\beta}(z), \\
{\cal S}^{(1)}_{\alpha\beta}(z)&=&
     \delta_{\alpha 1}\delta_{\beta 2}
      \theta(az_{1}-ay_{1})\theta(ay_{1}+b-az_{1})
      \theta(az_{2}-ay_{2})\theta(ay_{2}+b-az_{2}) \nonumber \\
    &&  \delta(az_{3}-ay_{3})\delta(az_{4}), \\
{\cal S}^{(2)}_{\alpha\beta}(z)&=&
      \delta_{\alpha 1}\delta_{\beta 2}
      \theta(az_{1}-ax_{1})\theta(ax_{1}+b-az_{1})
      \theta(az_{2}-ax_{2})\theta(ax_{2}+b-az_{2}) \nonumber \\
    &&  \delta(az_{3}-ax_{3})\delta(az_{4}-at).
\label{eqn:glue6}
\end{eqnarray}
This operator is shown diagrammatically in Figure~\ref{fig:glue}.

Substituting this into Eq.(\ref{eqn:glue4}), one finds in
the momentum representation
\begin{eqnarray}
\langle {\cal O}_1 \rangle_{m}^{cl}
  &=&  \exp\left\{
     -16 \pi^2\int\frac{d^4p}{(2\pi)^4}
     (e^{i\vec{p}\cdot\vec{y}} + e^{i\vec{p}\cdot\vec{x}+ip_4t})
     (e^{-i\vec{p}\cdot\vec{y}} + e^{-i\vec{p}\cdot\vec{x}-ip_4t})
           \right. \nn \\
  &&   \quad\quad \left. \times  
       \Pi_{j=1,2}\sbra{\frac{\sin(p_jb/2)}{p_j}}^2
       [\Delta D_0]^{-1}(p)\right\}. 
\label{eqn:glue7}
\end{eqnarray}

Since we study large $b$ behaviors, we use the
following formula
\begin{eqnarray}
\lim_{b\rightarrow\infty}
\left(\frac{\sin \alpha b}{\alpha}\right)^2
&=& \pi b \delta(\alpha).
\label{eqn:glue8}
\end{eqnarray}
Then we obtain
\begin{eqnarray}
\langle {\cal O}_1 \rangle_{m}^{cl}
&\simeq& \exp\left\{
-2\kappa\pi^2 b^2\int\frac{dp_3dp_4}{(2\pi)^2}
\sbra{\frac{1}{p_4^2+p_3^2+m_2^2} - \frac{1}{p_4^2+p_3^2+m_1^2}}
\right. \nn \\
  &&   \quad\quad \left. 
+ \kappa\pi^2 b^2\int\frac{dp_3}{2\pi}
\frac{e^{-E_{p_3}t}}{E_{p_3}} \cos p_3(x_3-y_3)
\right\},
\label{eqn:glue11}
\end{eqnarray}
where $E_{p_3}=\sqrt{p_3^2 + m_2^2}$. Since $m_1>>m_2$, we have neglected
the term proportional to $e^{-\sqrt{p_3^2 + m_1^2}t}$ in 
Eq.(\ref{eqn:glue11}).

Next the plaquette variable $S_{\alpha\beta}$ in Eq.(\ref{eqn:glue4})
for the $\langle {\cal O}_2 \rangle_{m}^{cl}$ is expressed by
\begin{eqnarray}
{\cal S}_{\alpha\beta}(z)&=& -{\cal S}^{(1)}_{\alpha\beta}(z)
                            +{\cal S}^{(2)}_{\alpha\beta}(z). 
\label{eqn:glue12}
\end{eqnarray}
The same calculation yields
\begin{eqnarray}
\langle {\cal O}_2 \rangle_{m}^{cl}
&\simeq& \exp\left\{
-2\kappa\pi^2 b^2\int\frac{dp_3dp_4}{(2\pi)^2}
\sbra{\frac{1}{p_4^2+p_3^2+m_2^2} - \frac{1}{p_4^2+p_3^2+m_1^2}}
\right. \nn \\
  &&   \quad\quad \left. 
- \kappa\pi^2 b^2\int\frac{dp_3}{2\pi}
\frac{e^{-E_{p_3}t}}{E_{p_3}} \cos p_3(x_3-y_3)
\right\}.
\label{eqn:glue13}
\end{eqnarray}

The plaquette variable $S_{\alpha\beta}$ in Eq.(\ref{eqn:glue4})
for the $\langle {\cal O}_3 \rangle_{m}^{cl}$ is 
${\cal S}^{(2)}_{\alpha\beta}(z)$ in Eq.(\ref{eqn:glue6})
and the result becomes
\begin{eqnarray}
\langle {\cal O}_3 \rangle_{m}^{cl}
&\simeq& \exp\left\{
-2\kappa\pi^2 b^2\int\frac{dp_3dp_4}{(2\pi)^2}
\sbra{\frac{1}{p_4^2+p_3^2+m_2^2} - \frac{1}{p_4^2+p_3^2+m_1^2}}
\right\}.
\label{eqn:glue15}
\end{eqnarray}

\begin{center}
\begin{figure}
\hspace{13mm}
\epsfig{file=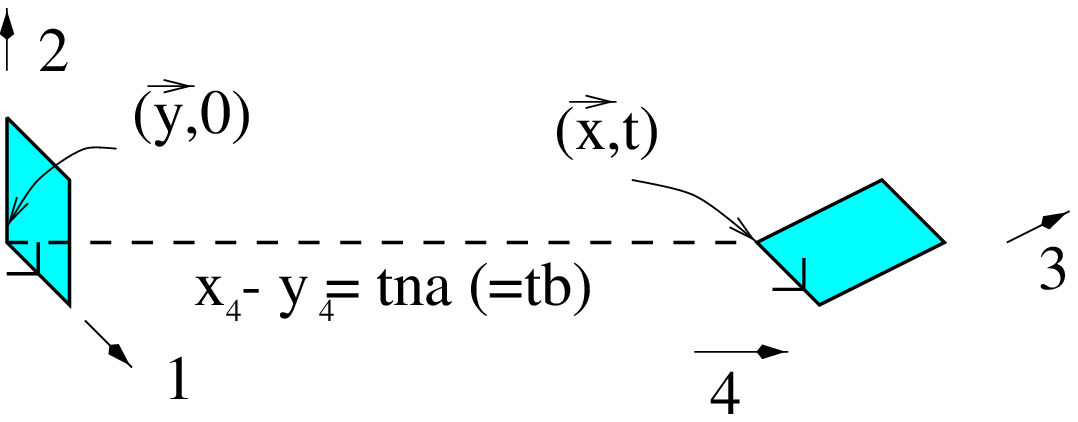,width=80mm}
\caption{
The plaquette variable $S_{\alpha\beta}$ for 
$\langle {\cal O}_4 \rangle_{m}^{cl}$.
}
\label{fig:gluebf}
\end{figure}
\end{center}

\begin{center}
\begin{figure}
\hspace{13mm}
\epsfig{file=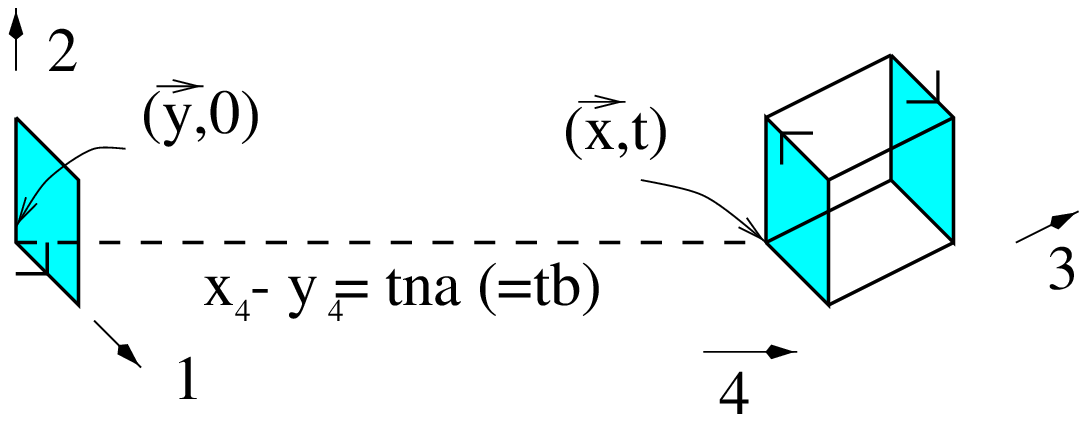,width=80mm}
\caption{
The plaquette variable $S_{\alpha\beta}$ for 
$\langle {\cal O}_4 \rangle_{m}^{cl}$. }
\label{fig:glueb}
\end{figure}
\end{center}

%%%%%%%%%%%%%%%%%%%%%%%%%%%%%%%%%%%%%%%%%%%%%%
For the opertor ${\cal O}_4$, a naive choice of $S_{\alpha\beta}$
in Figure~\ref{fig:gluebf} does not contribute. 
But when $S_{\alpha\beta}$ is chosen as in Figure~\ref{fig:glueb},
the classical part (\ref{eqn:glue4}) become non-zero and 
it is the leading contribution.
The plaquette variable $S_{\alpha\beta}$ in this case
is expressed by
%%%%%%%%%%%%%%%%%%%%%%%%%%%%%%%%%%%%%%%%%%%%%%
\begin{eqnarray}
{\cal S}_{\alpha\beta}(z)&=& {\cal S}^{(1)}_{\alpha\beta}(z)
                            +{\cal S}^{(3)}_{\alpha\beta}(z), \\
{\cal S}^{(3)}_{\alpha\beta}(z)&=&
      \delta_{\alpha 1}\delta_{\beta 2}
      \theta(az_{1}-ax_{1})\theta(ax_{1}+b-az_{1})
      \theta(az_{2}-ax_{2})\theta(ax_{2}+b-az_{2}) \nonumber \\
    &&  \sbra{ \delta(az_{3}-ax_{3}) - \delta(az_{3}-ax_{3}-b) 
         }\delta(az_{4}-at)\nonumber \\
    && + \delta_{\alpha 2}\delta_{\beta 3}
      \theta(az_{2}-ax_{2})\theta(ax_{2}+b-az_{2})
      \theta(az_{3}-ax_{3})\theta(ax_{3}+b-az_{3}) \nonumber \\
    &&  \sbra{ -\delta(az_{1}-ax_{1}) + \delta(az_{1}-ax_{1}-b) 
         }\delta(az_{4}-at)\nonumber \\
    && -  \delta_{\alpha 1}\delta_{\beta 3}
      \theta(az_{1}-ax_{1})\theta(ax_{1}+b-az_{1})
      \theta(az_{3}-ax_{3})\theta(ax_{3}+b-az_{3}) \nonumber \\
    &&   \delta(az_{2}-ax_{2}-b) \delta(az_{4}-at).
\label{eqn:glue6c}
\end{eqnarray}
This leads us to
\begin{eqnarray}
\langle {\cal O}_4 \rangle_{m}^{cl}
&\simeq& \exp\left\{
-\kappa\pi^2 b^2\int\frac{dp_3dp_4}{(2\pi)^2}
\mbra{5-4\cos{p_3x_3}}
\sbra{\frac{1}{p_4^2+p_3^2+m_2^2} - \frac{1}{p_4^2+p_3^2+m_1^2}}
\right. \nn \\
  &&   \quad\quad \left. 
+ \kappa\pi^2 b^2\int\frac{dp_3}{2\pi}
\frac{e^{-E_{p_3}t}}{E_{p_3}} \cdot
\frac{1}{2}\lbra{(1-e^{-ip_3x_3})e^{-ip_3(x_3-y_3)}+
                 (1-e^{ip_3x_3})e^{ip_3(x_3-y_3)}}
\right\}. \nonumber \\
\label{eqn:glue6d}
\end{eqnarray}

Finally, we get
\begin{eqnarray}
\langle\Psi (t)\cdot \Psi(0)\rangle_c 
&\simeq& \frac{6}{4V}\sum_{\vec{x},\vec{y}}
    \sbra{e^{-2A+B} + e^{-2A-B} -2e^{-2A}
     + 2e^{-A'+B'} + 2e^{-A'-B'} -4e^{-A'}},
\label{eqn:glue16}
\end{eqnarray}
where we define 
\begin{eqnarray}
 A &\equiv& \kappa\pi^2 b^2\int\frac{dp_3dp_4}{(2\pi)^2}
\sbra{\frac{1}{p_4^2+p_3^2+m_2^2} - \frac{1}{p_4^2+p_3^2+m_1^2}}
= \sigma_{cl}\cdot b^2,
\nonumber \\
\label{eqn:glue17}
 B &\equiv& \kappa\pi^2 b^2\int\frac{dp_3}{2\pi}
\frac{e^{-E_{p_3}t}}{E_{p_3}} \cos p_3(x_3-y_3),
\nonumber \\
\label{eqn:glue18}
A' &\equiv& \kappa\pi^2 b^2\int\frac{dp_3dp_4}{(2\pi)^2}
\mbra{5-4\cos{p_3x_3}}
\sbra{\frac{1}{p_4^2+p_3^2+m_2^2} - \frac{1}{p_4^2+p_3^2+m_1^2}},
\nonumber \\
\label{eqn:glue18a}
B' &\equiv& \kappa\pi^2 b^2\int\frac{dp_3}{2\pi}
\frac{e^{-E_{p_3}t}}{E_{p_3}}\cdot
\frac{1}{2}\lbra{(1-e^{-ip_3x_3})e^{-ip_3(x_3-y_3)}+
                 (1-e^{ip_3x_3})e^{ip_3(x_3-y_3)}}.
\label{eqn:glue18b}
\end{eqnarray}
Since $B$ and $B'$ contains $e^{-E_{p_3}t}$, it become very small
when $t >> 1$. Then one can expand the exponential and obtain finally
for $t >> 1$ 
\begin{eqnarray}
\langle\Psi (t)\cdot \Psi(0)\rangle_c 
%&=& \frac{6}{4V}e^A\sum_{\vec{x},\vec{y}}B^2 \nonumber \\
&\simeq& \frac{6}{4V}\sum_{\vec{x},\vec{y}}\sbra{e^{-A} B^2 + 2 e^{-A'}B'^2} 
%&=& \frac{6}{4}L^2e^A \int_{-\infty}^{\infty}dp_3
%= \frac{6}{4}L^2e^A \int_{-\infty}^{\infty}dp_3
%    \frac{e^{-2\sqrt{p_3^2 + m_2^2}t}}{p_3^2 + m_2^2}.
= \frac{6}{4}L^2e^{-2A} \int_{-\infty}^{\infty}\frac{dp_3}{2\pi}
    \frac{e^{-2\sqrt{p_3^2 + m_2^2}t}}{p_3^2 + m_2^2}
\nonumber \\
 \quad\quad\quad
  && + \frac{12}{4V}\sum_{\vec{x},\vec{y}}e^{-A'}
     \int_{-\infty}^{\infty}\frac{dp_3}{2\pi}
      2\mbra{1-\cos{p_3x_3}}
     \frac{e^{-2\sqrt{p_3^2 + m_2^2}t}}{p_3^2 + m_2^2}.
\label{eqn:glue19}
\end{eqnarray}
When $t >> 1$, the integrand decreases repidly and the 
integral is well approximated by the suddle point value at $p_3=0$.
Hence we get at large time $t$

\begin{eqnarray}
\langle\Psi (t)\cdot \Psi(0)\rangle_c 
\simeq \frac{3}{2\pi m_2^2} 
       \mbra{L^2e^{-2\sigma_{cl}\cdot b^2} + 
            2L(2L-1)e^{-5\sigma_{cl}\cdot b^2}}
       \exp\mbra{-2m_2 t},
\label{eqn:glue20}
\end{eqnarray}
where the second term coming from the ${\cal O}_4$ is seen to be
suppressed by the factor $e^{-3\sigma_{cl}\cdot b^2}$, since 
$\sigma_{cl}\cdot b^2$ become large for $b>>1$. Other quantum corrections
are also suppressed similarly.
The lowest glueball mass $M_{0^{++}}$ is found 
to be $M_{0^{++}}=2m_2$. 

The lowest glueball mass in unit of the string tension
$\sigma_{cl}$ for $b=$ 2.1, 2.9 and 3.8 are given in 
Table~\ref{tbl:mass}.
This is almost consistent with the recent lattice results
$M(0^{++})/ \sqrt{\sigma_{phys}} = 3.74\pm 0.12$ \cite{teper98}.

%%%%%%%%%%%%%%%%%%%%%%%%%%%%%%%%%%%%%%%%%%%%%%%%%%%%%%%%
%  Analysis of the results
%%%%%%%%%%%%%%%%%%%%%%%%%%%%%%%%%%%%%%%%%%%%%%%%%%%%%%%%
\section{Analysis}
\label{AN}

The value of the string tension calculated analytically in the previous
section is about two times larger than the value which is 
numerically determined from the monopole contribution to the 
abelian Wilson loop and is used here to fix the physical scale.

\begin{center}
\begin{figure}
\hspace{13mm}
\epsfig{file=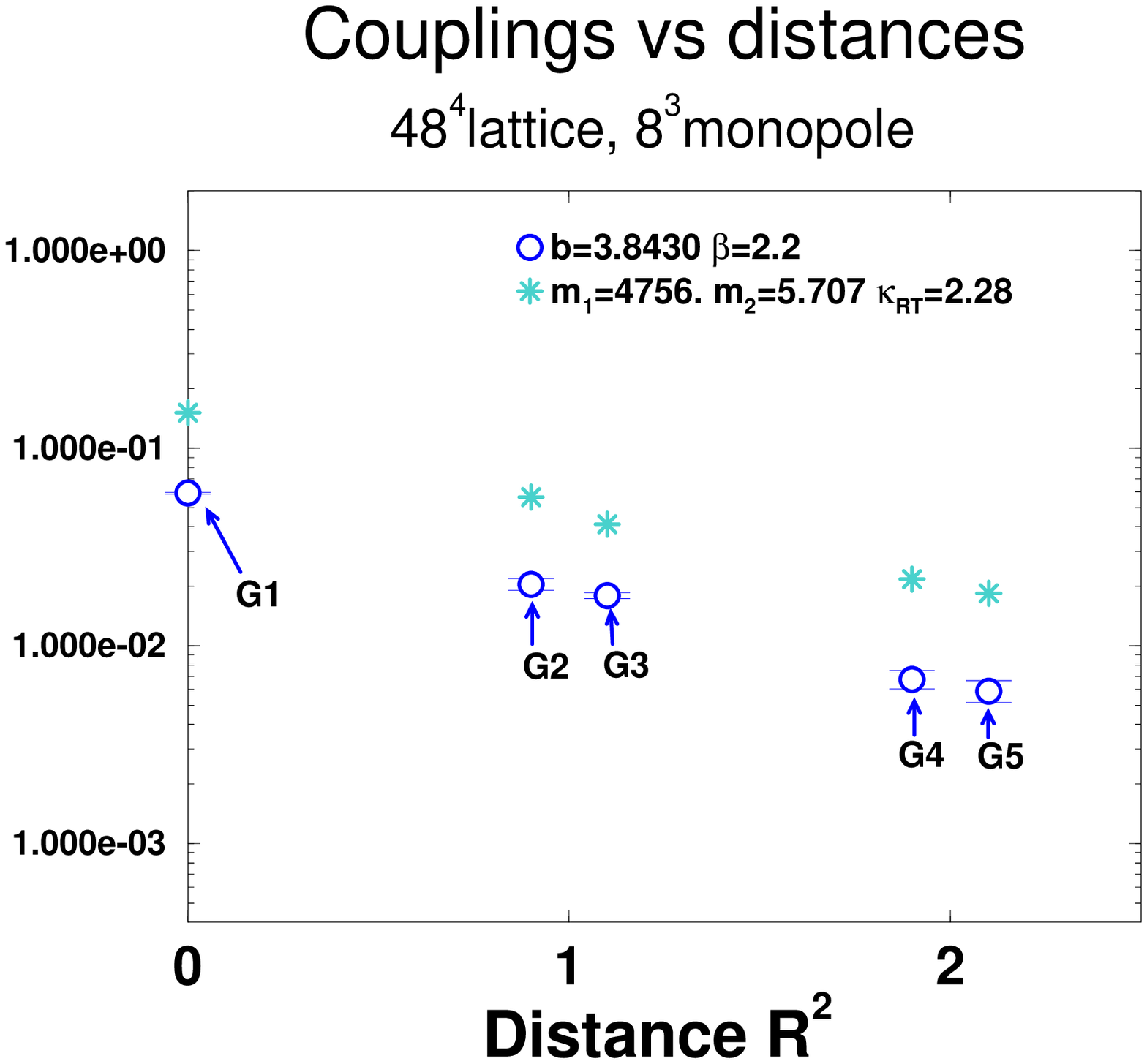,width=100mm}
\caption{
The expected coupling constants of RT (star) versus
numerical data. }
\label{fig:opt}
\end{figure}
\end{center}

Let us analyse the origin of the difference in details.
The method and the assumptions we have adopted are summarized in 
the following:
\begin{enumerate}
 \item {\bf Abelian dominance} We have assumed first that after abelian
       projection  abelian components alone are responsible for
       non-perturbative phenomena of $SU(2)$ Gluodynamics in the infrared
       region. This assumption is based on the numerical data obtained
       in MA gauge\cite{Kronfeld,domi}. Bali et al.\cite{bali96} have 
       made a
       detailed test at $\beta=2.5115$ and have confirmed the 
       assumption of abelian dominance of the string tension is good at
       the level of 92 percent. 
 \item {\bf Monopole dominance} The abelian Wilson operator can be
       factorized into monopole and photon contributions. We have
       assumed only the monopole part is responsible for the string
       tension on the basis of the numerical analysis
       \cite{shiba_suzuki,stack}. 
       The values of the string tension we have used are listed in Table 
       \ref{str}. The differences 
       are not big.
 \item {\bf DeGrand-Toussaint (DT) definition of lattice monopole} We have
       used DT monopole in the numerical evaluation done in Section
       ~\ref{PerfecAc}, since we do not know an alternative which can 
       be used in
       numerical simulations. The magnetic charge of DT
       monopole is restricted. However we have used the definition of
       lattice monopole with any integer charge 
       which we call as natural monopole
       in the step of the analytic
       block spin transformation. As checked in the case of compact
       QED\cite{shiba_suzuki},
       there may be a considerable difference 
       between natural and DT monopoles
       on the fine $a$ lattice for
       small $\beta$ region. But the difference is expected to be
       decreased after block spin transformations, since the blocked 
       monopole can take a wider range of charge. But we can not
       estimate the effect quantitatively in the present stage.   
 \item {\bf Truncation and scaling} In the inverse Monte-Carlo
       calculations and numerical block spin
       transformations, we have truncated the number of the terms in
       the effective monopole action. We have used 27 quadratic terms up 
       to 3 lattice distances and four-point  and six-point self
       interactions, assuming short-ranged interactions are more
       dominant. Then we have performed the block spin transformation 
       the number of steps of which is  $n=1,2,3,4,6,8$. The data seem
       to show roughly the scaling behavior expected on the renormalized
       trajectory. However, this step could still give rise to fairly large 
       systematic errors.
       The scaling behavior may not be enough. Actually, the dominant
       quadratic self-coupling term $G(1)$ at $b=2.78$ $(\beta=2.0, n=4)$
       is around 0.16, whereas it is around 0.09 at $b=2.87$
       $(\beta=2.3, n=8)$.
 \item {\bf Analytic calculations} Since the quadratic terms 
       seem to be dominant in the infrared region, we have evaluated the 
       physical quantities in the framework of the quadratic monopole action.
       Using the mean-field approximation, the quartic term can be
       approximated by the quadratic self and the nearest-neighbor terms 
       with an effective coupling $8q(b)<k_{\mu}^2(s)>$, where $q(b)$ is 
       the quartic coupling constant and $<k_{\mu}^2(s)>$ is the monopole
       density. The induced effective self-coupling is still by two or three
       order smaller than the original quadratic self-coupling.  Hence
       contributions from four and six point interactions  
       can be neglected safely for $b\ge 1.5 \sigma_{phys}^{-1/2}$.
       Since quantum corrections are also very small,
       we have made calculations using the classical
       contributions alone. The strong coupling expansion of the string
       model calculations is reliable.
       We show the expected coupling constants of RT for large $b$
       regions in Figure~\ref{fig:opt}. The comparison of the three
       parameters $\bar{\alpha}$ $\bar{\beta}$ $\bar{\gamma}$ between 
       the expected RT and the optimal fit to the numerical data are 
       plotted also in Table~\ref{tbl:opti}.
\end{enumerate}

As a results, we come to the conclusion that we have to perform 
Monte-Carlo simulations 
on an improved action for large $b$ starting from the points nearer 
to the continuum
and more steps of block spin
       transformations to reproduce the correct value of the string
       tension. It is stressed, however, that 
      the other parts of the above procedure appear rather reliable.

%%%%%%%%%%%%%%%%%%%%%%%%%%%%%%%%%%%%%%%%%%%%%%%%%%%%%%%%%%%%%%%%%%
%  Concluding remarks
%%%%%%%%%%%%%%%%%%%%%%%%%%%%%%%%%%%%%%%%%%%%%%%%%%%%%%%%%%%%%%%%%%
\section{Concluding remarks}
\label{concl}

\begin{enumerate}
\item In order to obtain the quantum perfect effective 
action of low-energy SU(2) Gluodynamics, we have performed 
the blockspin transformations on the dual lattice after
abelian projection in MA gauge numerically.
In the Inverse Monte-Carlo method, we have adopted more general
form of monopole actions than the one in the  
previous study \cite{shiba_suzuki,nakam}
and have stressed the important features of the almost perfect monopole 
action.
We have transformed the monopole action into that of 
the string model of hadrons by using the BKT transformation.

\item 
To evaluate the physical quantities, we have considered the 
quadratic interaction subspace for the monopole action and 
find the correct form of perfect operators.
We have evaluated the physical quantities
such as string tension and the glueball mass for SU(2) Gluodynamics 
using the string model of hadrons analytically. 
The strong coupling expansion works good and it turns out that 
the {\it classical} results in the string model is near to
the one in {\it quantum} $SU(2)$ Gluodynamics.
Probably, it means that the classical string
theory is a good approximation for IR Gluodynamics.
%In this paper we neglected the correction come from the higher order
%interactions for monopole currents. 

\item
To get a better fit of the string tension, we have to perform
more elaborate Monte-Carlo simulations for large $b$ 
on larger lattices.

\end{enumerate}

%%%%%%%%%%%%%%%%%%%%%%%%%%%%%%%%%%%%%%%%%%%%%%%%%%%%%%%%%%%%%%%%%%%
%
%%%%%%%%%%%%%%%%%%%%%%%%%%%%%%%%%%%%%%%%%%%%%%%%%%%%%%%%%%%%%%%%%%%
\section*{ACKNOWLEDGMENTS}
This work is supported 
by the Supercomputer Project (No.98-33, No.99-47)
of High Energy Accelerator Research Organization (KEK)
and the Supercomputer Project 
of the Institute of 
Physical and Chemical Research (RIKEN).
%-----------------------------------------------------------
T.S. acknowledges the financial support from  
JSPS Grant-in Aid for Scientific Research (B) (No.10440073 and
No.11695029).

%%%%%%%%%%%%%%%%%%%%%%%%%%%%%%%%%%%%%%%%%%%%%%
%\newpage

%%%%%%%%%%%%%%%%%%%%%%%%%%%%%%%%%%%%%%%%%%%%%%%%%%%%%%%%%%%%%%%%%%%
%     Appendix
%%%%%%%%%%%%%%%%%%%%%%%%%%%%%%%%%%%%%%%%%%%%%%%%%%%%%%%%%%%%%%%%%%%
%\newpage
\appendix
%%%%%%%%%%%%%%% Appendix A
\section{}\label{PerOper}
%%%%%%% quadratic couplings
%\input{p2_ap2}
%%%%%%%%%%%%%%%%%%%%%%%%%%%%%%%%%%%%%%%%%%%%%%%%%%%%%%%%%%
%     p2_ap2.tex   2000,2,7 modified
%%%%%%%%%%%%%%%%%%%%%%%%%%%%%%%%%%%%%%%%%%%%%%%%%%%%%%%%%

The quadratic interactions used for the modified Swendsen method
are shown in Table~\ref{tbl:appquad}.
Only the partner of the current multiplied by $k_{\mu}(s)$ are listed.
All terms in which the relation of the two currents is equivalent should
be added to satisfy translation and rotation invariances. 

The higher order interactions used for the modified Swendsen method
are listed in Table~\ref{tbl:higher}.

%%%%%%%%%%%%%%% Appendix B
\section{}\label{MoWil}
%%%%%%% MONOPOLE REPRESENTATIONS OF THE WILSON LOOP OPERATOR
%\input{p2_ap1}
%%%%%%%%%%%%%%%%%%%%%%%%%%%%%%%%%%%%%%%%%%%%%%%%%%%%%%%%%%%%
%        p2_ap1.tex  200,1,5 modified     
%%%%%%%%%%%%%%%%%%%%%%%%%%%%%%%%%%%%%%%%%%%%%%%%%%%%%%%%%%%%

In this appendix we give various representations of the
Wilson loop operator.

\begin{center}
{\bf The original representation}
\end{center}

Let us consider the generalized Villain action defined by 
Eq.(\ref{ap1:1}). In this model,
the quantum average of the Wilson loop operator 
is written as 
\begin{eqnarray}
\left< W(C) \right> &=&
  \left<
    \exp\left\{
      i \sum_{s,\mu} J_{\mu}(s)\theta_{\mu}(s) 
    \right\}
  \right>
= Z[J]/Z[0], \\
Z[J] &\equiv&
  \int_{-\pi}^{\pi}\prod_{s;\mu}d\theta_{\mu}(s)
  \sum_{n_{\mu\nu}(s)=-\infty}^{+\infty}
  \exp\left\{
    -{\cal S}[\theta,n] + i \sum_{s,\mu}J_{\mu}(s)\theta_{\mu}(s)
  \right\}.
\label{ap1.1}
\end{eqnarray}
We call this as the original representation of the Wilson loop.

\begin{center}
{\bf The monopole representation}
\end{center}
The above original representation can be transformed into the
monopole representation exactly in the following way.
Let us perform the BKT-transformation with respect to the 
integer-valued tensor $n_{\mu\nu}(s)$ in (\ref{ap1.1}):
\begin{eqnarray}
 n_{\mu\nu}(s) &=& m_{\mu\nu}(s) + \partial_{[\mu}q_{\nu]}(s), \\
 \partial_{[\mu}m_{\nu\rho]}(s) &\equiv& 
      \frac{1}{2}\epsilon_{\mu\nu\rho\lambda}k_{\lambda}
                 (s-\hat{\mu}-\hat{\nu}-\hat{\rho}),
\label{ap1.2}
\end{eqnarray}
where $m_{\mu\nu}(s)$ and $q_{\mu}(s)$ are rank-2 tensor and 
vector fields on the original lattice respectively.
The vector field $k_{\mu}(s)$ which can be interpreted as a 
monopole current on the dual lattice obeys conservation law
$\partial_{\mu}^{'}k_{\mu}(s)=0$ by definition.

Using the Hodge-de-Rahm decomposition we write
\begin{eqnarray}
\partial_{[\mu}\theta_{\nu]}(s)+2\pi n_{\mu\nu}(s) &=&
  \partial_{[\mu}\theta^{(n.c)}_{\nu]}(s)
  +2\pi \sum_{s'}
  \partial'_\rho \Delta_L^{-1}(s-s') 
      \frac{1}{2}\epsilon_{\rho\mu\nu\lambda}k_{\lambda}
                 (s'-\hat{\rho}-\hat{\mu}-\hat{\nu})
\label{ap1.3}
\\
\theta^{(n.c)}_{\mu}(s) &=&
  \theta_{\mu}(s)
  +2\pi \sum_{s'}
  \Delta_L^{-1}(s-s') \partial'_\nu m_{\mu\nu}(s') +q_{\mu}(s).
\label{ap1.4}
\end{eqnarray}
Substituting Eq.(\ref{ap1.3}) in Eq.(\ref{ap1.1}) and integrating out
the noncompact field $\theta^{(n.c)}_{\mu}(s)$ we get 
\begin{eqnarray}
Z[J] &=&
  \sum_{k_{\mu}(s)=-\infty}^{+\infty}
  \sum_{m_{\mu\nu}(s)=-\infty}^{+\infty}
  \!\!\!\exp\Biggl\{
    -\pi^2 \sum_{s,s';\mu}
    J_\mu(s)\left( \frac{1}{\Delta_L^2 D_0} \right)(s-s')J_\mu(s')
  \Biggr.
\nonumber \\
&&\qquad\qquad
  \Biggl.
    -\sum_{s,s';\mu}
    k_{\mu}(s) D_0(s-s') k_{\mu}(s')
    -2\pi i\sum_{s,s'}
    J_\mu(s) \Delta_L^{-1}(s-s') \partial'_\nu m_{\mu\nu}(s')
  \Biggr\}.
\label{ap1.5}
\end{eqnarray}
It is convenient to define the plaquette variable $S^J_{\beta\gamma}(s)$
from the abelian integer-charged electric current $J_{\gamma}(s)$
by the following relation
\begin{eqnarray}
\partial'_{\beta}S^J_{\beta\gamma}(s)=J_{\gamma}(s).
\end{eqnarray}
By this definition, $S_{\mu\nu}(s)$ can be interpreted as 
the surface which is spanned on the contour $J_{\gamma}(s)$.
The third term on the exponential in Eq.(\ref{ap1.5}) can be 
rewritten as follows:
\begin{eqnarray}
\sum_{s,\mu} J_\mu(s) \Delta_L^{-1} \partial'_\nu m_{\mu\nu}(s)
&=&
  \sum_{s,\mu} \partial'_\rho S^J_{\mu\rho}(s)
  \Delta_L^{-1} \partial'_\nu m_{\mu\nu}(s)
\nonumber\\
&=&
  \sum_{s,\mu} S^J_{\mu\rho}(s)
  \partial_{\rho}\Delta_L^{-1} \partial'_\nu m_{\mu\nu}(s)
\nonumber \\
&=&
  \sum_{s,\mu} S^J_{\mu\rho}(s)
  m_{\mu\rho}(s)
  - \sum_{s,\mu} S^J_{\mu\rho}(s)
  \partial'_{\nu}\Delta_L^{-1}\partial_{[\rho}m_{\mu\nu]}(s).
\end{eqnarray}
When use is made of Eq.(\ref{ap1.2}), we have
\begin{eqnarray}
\sum_{s,\mu} S^J_{\mu\rho}(s)
  \partial'_{\nu}\Delta_L^{-1}\partial_{[\rho}m_{\mu\nu]}(s)
&=& \sum_{s,s'} k_{\mu}(s) \Delta_L^{-1}(s-s')\frac{1}{2}
  \epsilon_{\mu\alpha\beta\gamma}\partial_{\alpha}
  S^J_{\beta\gamma}(s'+\hat{\mu})  \nonumber \\
&=& \sum_{s,\mu}N_{\mu}(s)k_{\mu}(s),
\end{eqnarray}
where $N_{\mu}(s)$ is defined by (\ref{ap1?4}).

The summation with respect to the integer field $m_{\mu\nu}(s)$ is 
trivial since $\exp\mbra{2\pi i \times \mbox{integer}}=1$.
%--------------------------------------------
Therefore, the expectation value of the Wilson loop operator in the 
monopole representation becomes:
\begin{eqnarray}
\left< W(C) \right> =
  \left< W(C) \right>_m \cdot 
  \exp\left\{
    -\pi^2 \sum_{s,s'}\sum_{\mu} J_{\mu}(s)
    \left( \frac{1}{\Delta_L^2 D_0} \right)(s-s')J_{\mu}(s')
  \right\},
\label{eq:2-1}
\end{eqnarray}
where $\left< W(C) \right>_m$ is written as
\begin{eqnarray}
\left< W(C) \right>_m &=&
  \left<
    \exp\left\{
      2\pi i \sum_{s,\mu}k_{\mu}(s)N_{\mu}(s)
          \right\}
  \right> = Z[J]/Z[0], \\
Z[J]  &=& \sum_{k_{\mu}(s)=-\infty}^{\infty}
  \left( \prod_s \delta_{\partial_{\mu}'k_{\mu}(s),0} \right)
  \exp\left\{
    -{\cal S}[k] + 2\pi i \sum_{s,\mu}k_{\mu}(s)N_{\mu}(s)
  \right\}.
\label{eq:2-2}     
\end{eqnarray}
The monopole action ${\cal S}[k]$ is shown in (\ref{ap1?3}).

{\it Note that the difference between $\left< W(C) \right>_m$ and
$\left< W(C) \right>$
is only an electric-electric current $J-J$ interaction which  
comes from the exchange of regular photons and has no line 
singularity leading to a linear potential. Hence the term of the area
law of both operators are completely the same.}
So concerning the
low-energy physics of QCD, such a term is not so important. We,
therefore, neglect $J-J$ interactions and consider
$\left< W(C) \right>_m$
to evaluate the static potential. 
The analysis in \cite{ourpaper} leads to Eq.(\ref{opwil:5}).

\begin{center}
{\bf The dual representation}
\end{center}
%------------ monopole -> DAH ------------

As is written in \cite{SvdS} the theory described by the monopole
action (\ref{ap1?3}) is given in the particle representation.
It can be expressed in the field representation as a field theory.
This is a dual abelian Higgs model.
We show here the above monopole representation is equivalent to 
the lattice form of the modified London limit of the dual 
abelian Higgs model.

Introducing an auxiliary dual field $\phi(s)$ for the constraint of 
the monopole current $\delta_{\partial_{\mu}'k_{\mu}(s),0}$ and 
a dual vector field $C_{\mu}(s)$,
Eq.(\ref{eq:2-2}) is rewritten as
\begin{eqnarray}
Z[J] &=&
  \exp\left\{
    -\pi^2 \sum_{s,s';\mu}
    J_\mu(s)\left( \frac{1}{\Delta_L^2 D_0} \right)(s-s')J_\mu(s')
  \right\}
\nonumber \\
&&
  \times
  \int_{-\infty}^{\infty}\prod_{s;\mu}d C_{\mu}(s)
  \int_{-\pi}^{\pi}\prod_{s;\mu}d \phi(s)
  \sum_{k_{\mu}(s)={-\infty}}^{\infty}
  \exp\Biggl\{
    -\frac{1}{4\beta}
    \sum_{s;\mu>\nu}\left( \partial'_{[\mu}C_{\nu]}(s) \right)^2
  \Biggr.
\nonumber \\
&&\qquad\qquad
  \Biggl.
    +i\sum_{s;\mu}
    \left( C_\mu(s) + \partial'_\mu\phi(s) - 2\pi N_\mu(s) \right)
    k_\mu(s)
    -\sum_{s,s';\mu}k_\mu(s) D_0^{'-1}(s-s')k_\mu(s')
  \Biggr\}. \nonumber \\
\label{eq:2c-1}
\end{eqnarray}
Inserting the unity
$1=\int_{-\infty}^{\infty}{\cal D}F\delta\sbra{F_{\mu}(s)-k_{\mu}(s)}$ 
to Eq. (\ref{eq:2c-1})
and performing the Gaussian integration with respect to the 
$F_\mu(s)$ field, we have 
\begin{eqnarray}
Z[J] &=&
  \exp\left\{
    -\pi^2 \sum_{s,s';\mu}
    J_\mu(s)
    \left(
       \frac{1}{\beta\Delta_L} -  \frac{1}{\Delta_L^2 D_0}
    \right)(s-s')J_\mu(s')
  \right\}
\nonumber \\
&&
  \int_{-\infty}^{\infty}\prod_{s;\mu}d C_{\mu}(s)
  \int_{-\pi}^{\pi}\prod_{s;\mu}d \phi(s)
  \sum_{l_{\mu}(s)={-\infty}}^{\infty}
%\nonumber \\
%&&
%  \times
  \exp\Biggl\{
    -\frac{1}{4\beta}
    \sum_{s;\mu>\nu}
    \left( \partial'_{[\mu}C_{\nu]}(s) - 2\pi S^J_{\mu\nu}(s) \right)^2
  \Biggr.
\nonumber \\
&&
  \Biggl.
    -\frac{1}{4}\sum_{s,s';\mu}
    \left(
      C_\mu(s) + \partial'_\mu\phi(s) + 2\pi l_\mu(s)
    \right) D_0^{'-1}(s-s')
    \left(
      C_\mu(s') + \partial'_\mu\phi(s') + 2\pi l_\mu(s')
    \right)
  \Biggr\}, \nonumber \\
\end{eqnarray}
where we have used also the Poisson summation formula
\begin{eqnarray}
 \sum_{k_{\mu}(s)={-\infty}}^{\infty} \delta\sbra{F_{\mu}(s)-k_{\mu}(s)}
  = \sum_{l_{\mu}(s)={-\infty}}^{\infty}
      \exp\mbra{2\pi i \sum_{s,\mu} F_{\mu}(s)l_{\mu}(s)}.
\label{eq:2c-2}
\end{eqnarray}
Therefore the expectation value of the Wilson loop operator in the
dual representation becomes
\begin{eqnarray}
\left< W(C) \right> &=&   \exp\left\{
    -\pi^2 \sum_{s,s';\mu}
    J_\mu(s)
    \left(
       \frac{1}{\beta\Delta_L} -  \frac{1}{\Delta_L^2 D_0}
    \right)(s-s')J_\mu(s')
  \right\}  \left< H({\cal C})\right>,
\label{eq:2c-4}
\end{eqnarray}
where $H({\cal C})$ is a 't Hooft loop operator defined by 
Eq.(\ref{ap1:?2}). We see
\begin{eqnarray}
 \left< H({\cal C})\right>
 &=& {\cal Z}[S_J]/{\cal Z}[0],
\nonumber \\
{\cal Z}[S_J] &=&   
  \int_{-\infty}^{\infty}\prod_{s;\mu}d C_{\mu}(s)
  \int_{-\pi}^{\pi}\prod_{s;\mu}d \phi(s)
  \sum_{l_{\mu}(s)={-\infty}}^{\infty}
  \exp\Biggl\{
    -\frac{1}{4\beta}
    \sum_{s;\mu>\nu}
    \left( \partial'_{[\mu}C_{\nu]}(s) - 2\pi S^J_{\mu\nu}(s) \right)^2
  \Biggr.
\nonumber \\
&&
  \Biggl.
    -\frac{1}{4}\sum_{s,s';\mu}
    \left(
      C_\mu(s) + \partial'_\mu\phi(s) + 2\pi l_\mu(s)
    \right) D_0^{'-1}(s-s')
    \left(
      C_\mu(s') + \partial'_\mu\phi(s') + 2\pi l_\mu(s')
    \right)
  \Biggr\}. \nonumber \\
\label{eq:2c-5}
\end{eqnarray}
Eq.(\ref{ap1:?}) is
the lattice form of the modified London limit of the dual 
abelian Higgs model. $C_\mu(s)$ and $\phi(s)$ can be interpreted as
a dual abelian gauge field and the phase variable of the dual Higgs 
field, respectively. 
Note that the integer-valued field $l_\mu(s)$ appears
due to the compactness of the theory.

%---------------------------------------
\begin{center}
{\bf The string representation}
\end{center}

We show here the string representation is obtained from the 
monopole representation.
Introducing an auxiliary field $\phi(s)$ for the constraint of 
the monopole current $\delta_{\partial_{\mu}'k_{\mu}(s),0}$ and
inserting the unity
$1=\int_{-\infty}^{\infty}{\cal D}F\delta\sbra{F_{\mu}(s)-k_{\mu}(s)}$ 
into Eq.(\ref{eq:2-2}), it is rewritten as
\begin{eqnarray}
Z[J] &=&
  \exp\left\{
    -\pi^2 \sum_{s,s';\mu}
    J_\mu(s)\left( \frac{1}{\Delta_L^2 D_0} \right)(s-s')J_\mu(s')
  \right\}
\nonumber \\
&&
  \times
  \int_{-\infty}^{+\infty}{\cal D}F_{\mu}(s)
  \int_{-\pi}^{\pi}\prod_{s}d\phi(s)
  \sum_{l_{\mu}(s)=-\infty}^{\infty}
  \exp\left\{
    -\frac{1}{4}\sum_{s,s';\mu}F_{\mu}(s)D_0(s-s')
      F_{\mu}(s')\right.
\nonumber \\
&&  \left.
    +i\sum_{s,\mu}F_\mu(s)
    \left(
      \partial'_\mu\phi(s) + 2\pi l_\mu(s) + 2\pi N_\mu(s)
    \right)
  \right\}. \nonumber \\
%\label{eq:2-d1}
\end{eqnarray}
Here we also have used the Poisson summation formula for the 
integer valued vector field $k_{\mu}(s)$.

Now we perform the BKT transformation with respect to the 
integer valued vector field $l_\mu(s)$: 
\begin{eqnarray}
l_{\mu}(s)= s_{\mu}(s) + \partial_{\mu}r(s),
\quad\quad \partial_{[ \mu}s_{\nu ]}(s) \equiv 
   \frac{1}{2}\epsilon_{\mu\nu\alpha\beta}
        \sigma_{\alpha\beta}(s-\hat{\alpha}-\hat{\beta}),
%\label{eq:2-d1}
\end{eqnarray}
where $r(s)$ is a scalar field defined on the dual lattice
and the string field $\sigma_{\alpha\beta}(s)$ defined on the original
lattice obeys the conservation law 
$\partial'_{\mu}\sigma_{\mu\nu}(s) =0$ by definition.
This means the variables $\sigma_{\alpha\beta}(s)$
form a closed surface on the original lattice. 

Integrating out all fields except for the string field 
$\sigma_{\alpha\beta}(s)$
, we obtain the following string representation defined
on the original lattice:
\begin{eqnarray}
Z[J] &=&
  \exp\left\{
    -\pi^2 \sum_{s,s';\mu}
    J_\mu(s)\left( \frac{1}{\Delta_L^2 D_0} \right)(s-s')J_\mu(s')
    -\pi^2 \sum_{s,s';\mu}
    N_\mu(s)\left( \frac{1}{D_0} \right)(s-s')N_\mu(s')
  \right\}
\nonumber \\
&&
  \times
  \sum_{\sigma_{\mu\nu}(s)=-\infty}^{\infty}
  \left(
    \prod_s \delta_{\partial'_{\mu}\sigma_{\mu\nu}(s),0}
  \right)
  \exp\Biggl\{
    -\pi^2\sum_{s,s';\mu>\nu}\sigma_{\mu\nu}(s)
    \left( \frac{1}{\Delta_L D_0} \right)(s-s')\sigma_{\mu\nu}(s')
  \Biggr.
\nonumber \\
&&\qquad\qquad\qquad\qquad
  \Biggl.
    -2\pi^2\sum_{s,s';\mu>\nu}
     \frac{1}{2}\epsilon_{\mu\nu\alpha\beta}
                 \sigma_{\alpha\beta}(s-\hat{\mu}-\hat{\nu})
    \left( \frac{1}{\Delta_L D_0} \right)(s-s')
       \partial'_{[\mu}N_{\nu]}(s')
  \Biggr\}. \nonumber \\
\end{eqnarray}

%%%%%%%%%%%%%%%%%%%%%%%%%%%%%%%%%%%%%%%%%%%%%%%%%%%%%%%%%%%%%%%%
\newpage

%\vspace*{.3cm}
%------------ fig 1 -----------------------------

%------------- fig 2 -----------------------------

%------------ fig 3 ---------------------------------

\vspace{1cm}
%----------- fig 4 ------------------------------------

%----------------- fig 5 --------------------------------
\vspace{1cm}

%---------------- fig 6 ---------------------------

\newpage
%----------------- fig 7a ---------------------------------

%----------------- fig 7b ---------------------------------

%----------------- fig 7c ---------------------------------

\newpage
%----------------- fig 8 --------------------------------

%%%%%%%%%%%%%%%%%%%%%%%%%%%%%%%%%%%%%%%%%%%%%%%%%%%%%%%%%%%%%%%%%%%%%%%%%
\newpage

\begin{table}
\caption{
The optimal values $\kappa$, $m_1$ and $m_2$
for $b=$ 2.1, 2.9 and 3.8
from the Inverse Monte-Carlo method}
\label{tbl:fit} 
\begin{tabular}[t]{clll}
{$b$}     & 2.1  & 2.9  & 3.8  \\ \hline
{$\kappa$}  & 1.76 & 3.12 & 4.83 \\
{$m_1$}     & 1.0 {$\times 10^4$} & 1.0 {$\times 10^4$} 
& 1.0 {$\times 10^4$} \\
{$m_2$}     & 12.0 & 12.0 & 12.0 \\
\end{tabular}
\end{table}

\begin{table}
\caption{
$\sqrt{\sigma_{cl}/\sigma_{phys}}$ for $b=$ 2.1, 2.9 and 3.8.}
\label{tbl:str} 
\begin{tabular}{clll}
$b$       & 2.1  & 2.9  & 3.8  \\ \hline
$\sqrt{\frac{\sigma_{cl}}{\sigma_{phys}}}$
          & 1.64 & 1.56 & 1.45 \\
\end{tabular}
\end{table}

\begin{table}
\caption{
The leading quantum correction for $b=$ 2.1, 2.9 and 3.8.}
\label{tbl:corr} 
\begin{tabular}{clll}
$b$       & 2.1 & 2.9 & 3.8 \\ \hline
$\frac{4}{b^2}e^{-4\Pi(0)b^2}$
          & $1.26\times 10^{-5}$ & $1.40\times 10^{-9}$
          & $1.65\times 10^{-14}$ \\
\end{tabular}
\end{table}

\begin{table}
\caption{
$M_{0^{++}}/\sqrt{ \sigma_{cl}}$ for $b=$ 2.1, 2.9 and 3.8.}
\label{tbl:mass} 
\begin{tabular}{clll}
$b$       & 2.1  & 2.9  & 3.8  \\ \hline
$M_{0^{++}}/\sqrt{ \sigma_{cl}}$
          & 5.56 & 4.18 & 3.36 \\
\end{tabular}
\end{table}

\begin{table}
\caption{
$\sqrt{\sigma_{L}/\sigma_{phys}}$ for $b=$ 2.1, 2.9 and 3.8
from Eq.(\ref{on-axis}).}
\label{tbl:str_L} 
\begin{tabular}{clll}
$b$       & 2.1  & 2.9  & 3.8  \\ \hline
$\sqrt{\frac{\sigma_{L}}{\sigma_{phys}}}$
          & 1.73 & 1.59 & 1.39 \\
\end{tabular}
\end{table}

\begin{table}
\caption{String tensions from non-abelian $(\sigma_f)$
%\cite{JF}
and monopole $(\sigma_m)$ Wilson loops.}
\label{str} 
\begin{tabular}{cll}
$\beta$ & $\sqrt{\sigma_f} a$ & $\sqrt{\sigma_m} a$ \\ \hline
2.20    & 0.4690(100)         & 0.4804(52)          \\
2.30    & 0.3690(30)          & 0.3589(36)          \\
2.40    & 0.2660(20)          & 0.2678(82)          \\
2.50    & 0.1905(8)           & 0.1851(32)          \\
2.60    & 0.1360(40)          & 0.1346(39)          \\
2.70    & 0.1015(10)          & 0.1016(21)          \\
\end{tabular}
\end{table}

\begin{table}
\caption{The comparison of the three
parameters $\bar{\alpha}$ $\bar{\beta}$ $\bar{\gamma}$ between 
the expected RT and the optimal fit to the numerical data.}
\label{tbl:opti} 
\begin{tabular}{clll}
$b$                  & 2.1                  & 2.9                  & 3.8   \\
\hline
$\bar{\alpha}$       & 0.565                & 0.321                & 0.207 \\
$\bar{\beta}$        & 6.78                 & 3.85                 & 2.49  \\
$\bar{\gamma}$       & $5.65\times 10^{-5}$ & $3.21\times 10^{-5}$ & $2.07\times 10^{-5}$ \\
\hline
$\bar{\alpha}_{RT}$  & 1.52                 & 0.780                & 0.435 \\
$\bar{\beta}_{RT}$   & 6.78                 & 3.85                 & 2.49  \\
$\bar{\gamma}_{RT}$  & $4.09\times 10^{-4}$ & $1.90\times 10^{-4}$ & $9.15\times 10^{-5}$ \\
\end{tabular}
\end{table}

%-----------------------------------

\begin{table}
\caption{The quadratic interactions used for the modified Swendsen method.}
\label{tbl:appquad} 
\begin{tabular}{cllcll}
{\it coupling $\mbra{G_i}$} &  distance& $ \ \ \ \ \ \ \ \ $  {\it type} $ \ \ \ \ \ \ \ \ $ &
{\it coupling $\mbra{G_i}$} &  distance& $ \ \ \ \ \ \ \ \ $  {\it type} \\ 
\hline
$G_1$    & (0,0,0,0) & $k_\mu(s)$ &
$G_{15}$ & (2,1,1,0) & $k_\mu(s+2\hat{\mu}+\hat{\nu}+\hat{\rho})$ \\
$G_2$    & (1,0,0,0) & $k_\mu(s+\hat{\mu})$ &
$G_{16}$ & (1,2,1,0) & $k_\mu(s+\hat{\mu}+2\hat{\nu}+\hat{\rho})$ \\
$G_3$    & (0,1,0,0) & $k_\mu(s+\hat{\nu})$ &
$G_{17}$ & (0,2,1,1) & $k_\mu(s+2\hat{\nu}+\hat{\rho}+\hat{\sigma})$ \\
$G_4$    & (1,1,0,0) & $k_\mu(s+\hat{\mu}+\hat{\nu})$ &
$G_{18}$ & (2,1,1,1) & $k_\mu(s+2\hat{\mu}+\hat{\nu}+\hat{\rho}+\hat{\sigma})$ \\
$G_5$    & (0,1,1,0) & $k_\mu(s+\hat{\nu}+\hat{\rho})$ &
$G_{19}$ & (1,2,1,1) & $k_\mu(s+\hat{\mu}+2\hat{\nu}+\hat{\rho}+\hat{\sigma})$ \\
$G_6$    & (2,0,0,0) & $k_\mu(s+2\hat{\mu})$ &
$G_{20}$ & (2,2,0,0) & $k_\mu(s+2\hat{\mu}+2\hat{\nu})$ \\
$G_7$    & (0,2,0,0) & $k_\mu(s+2\hat{\nu})$ &
$G_{21}$ & (0,2,2,0) & $k_\mu(s+2\hat{\nu}+2\hat{\rho})$ \\
$G_8$    & (1,1,1,1) & $k_\mu(s+\hat{\mu}+\hat{\nu}+\hat{\rho}+\hat{\sigma})$ &
$G_{22}$ & (3,0,0,0) & $k_\mu(s+3\hat{\mu})$ \\
$G_9$    & (1,1,1,0) & $k_\mu(s+\hat{\mu}+\hat{\nu}+\hat{\rho})$ &
$G_{23}$ & (0,3,0,0) & $k_\mu(s+3\hat{\nu})$ \\
$G_{10}$ & (0,1,1,1) & $k_\mu(s+\hat{\nu}+\hat{\rho}+\hat{\sigma})$ &
$G_{24}$ & (2,2,1,0) & $k_\mu(s+2\hat{\mu}+2\hat{\nu}+\hat{\rho})$ \\
$G_{11}$ & (2,1,0,0) & $k_\mu(s+2\hat{\mu}+\hat{\nu})$ &
$G_{25}$ & (1,2,2,0) & $k_\mu(s+\hat{\mu}+2\hat{\nu}+2\hat{\rho})$ \\ 
$G_{12}$ & (1,2,0,0) & $k_\mu(s+\hat{\mu}+2\hat{\nu})$ &
$G_{26}$ & (0,2,2,1) & $k_\mu(s+2\hat{\nu}+2\hat{\rho}+\hat{\sigma})$ \\
$G_{13}$ & (0,2,1,0) & $k_\mu(s+2\hat{\nu}+\hat{\rho})$ &
$G_{27}$ & (2,1,1,0) & $k_\rho(s+2\hat{\mu}+2\hat{\nu}+\hat{\rho})$ \\
$G_{14}$ & (2,1,0,0) & $k_\nu(s+2\hat{\mu}+\hat{\nu})$ &
         &           & \\
\end{tabular}
\end{table}

\begin{table}
\caption{The higher order interactions used for
the modified Swendsen method.}
\label{tbl:higher} 
\begin{tabular}{cll}
{\it coupling } & distance  & $ \ \ \ \ \ \ \ \ $  {\it type} \\ 
\hline
4-point         & (0,0,0,0) &
$\quad \quad \sum_{s}\sbra{\sum_{\mu=-4}^4 k_\mu^2(s)}^2$ \\
6-point         & (0,0,0,0) &
$\quad \quad \sum_{s}\sbra{\sum_{\mu=-4}^4 k_\mu^2(s)}^3$ \\
\end{tabular}
\end{table}


\begin{references}
%1
\bibitem{Wilson} K.G.~Wilson, {\it Phys.~Rev.} {\bf D14} (1974) 2455.
%2
\bibitem{'thooft} G.'t~Hooft, {\it Nucl.~Phys.} {\bf B190} (1981)
455.
%3
\bibitem{Abrikosov} A.A.~Abrikosov, {\it JETP} {\bf 32} (1957) 1442.
%4
\bibitem{Kronfeld} A.S.~Kronfeld {\it et al.}, {\it Phys.~Lett.} {\bf
198B} (1987) 516; A.S.~Kronfeld, G.~Schierholz and U.J. Wiese, {\it
Nucl.~Phys.} {\bf B293} (1987) 461.
%5
\bibitem{domi} T.~Suzuki and I.~Yotsuyanagi, {\it Phys. Rev.}
{\bf D42} (1990) 4257; {\it Nucl.~Phys.} {\bf B} {\it (Proc.~Suppl.)}
{\bf 20} (1991) 236; S.~Hioki {\it et al.}, {\it Phys.~Lett.} {\bf
B272} (1991) 326 and references therein.
%6
\bibitem{Reviews} T.~Suzuki, {\it Nucl.~Phys.} {\bf B} {\it
(Proc.~Suppl.)} {\bf 30} (1993) 176; \\ M.~I.~Polikarpov, {\it
Nucl.~Phys.} {\bf B} {\it (Proc.~Suppl.)} {\bf 53} (1997) 134; \\
G.~S.~Bali, talk given at 3rd
International Conference on Quark Confinement and the Hadron Spectrum
(Confinement III), Newport News, VA, June 1998, {\tt hep-ph/9809351}; \\
R.~W.~Haymaker, to be published in {\it Phys.~Rept.}, {\tt
hep-lat/9809094}.
%7
\bibitem{ChPoRev} M.N.~Chernodub and M.I.~Polikarpov, in {\it "Confinement,
Duality and Nonperturbative Aspects of QCD"}, p.387, Ed. by Pierre van Baal,
Plenum Press, 1998; {\tt hep-th/9710205}
%8
\bibitem{shiba_suzuki} H.~Shiba and T.~Suzuki,
{\it Phys.~Lett.} {\bf B343} (1995) 315,
{\it Phys.~Lett.} {\bf B351} (1995) 519
and references therein.
%9
\bibitem{Swendsen} R.H.~Swendsen, {\it Phys.~Rev. ~Lett.} {\bf 52}
(1984) 1165; {\it Phys.~Rev.} {\bf B30} (1984) 3866,3875.
%10
\bibitem{SvdS} J.~Smit and A.J.~van~der~Sijs, {\it Nucl. Phys.} {\bf
B355} (1991) 603.
%11
\bibitem{nakam} S.~Kato, S.~Kitahara, N.~Nakamura and T.~Suzuki,
{\it Nucl.~Phys.} {\bf B520} (1998) 323.
%12
\bibitem{kato98} S. Fujimoto et al, Nucl.Phys.B(Proc.Suppl)73(1999) 533.
%13
\bibitem{D_T} T.A. DeGrand and D. Toussaint, Phys. Rev. 
{\bf D22} (1980) 2478.
%14
\bibitem{ivanenko} T.L. Ivanenko, A.V. Pochinskii and
M.I. Polikarpov, {\it Phys. Lett.} {\bf B 252} (1990) 631.
%15
\bibitem{ourpaper} S. Fujimoto et al, A Quantum Perfect Lattice Action
 for Monopoles and Strings, to be published in {\it Phys. Lett.} {\bf B},
{\tt hep-lat/0002006}.
%16
\bibitem{suzu89} T. Suzuki, Prog.Theor.Phys. {\bf 80} (1988) 929; {\bf
81} (1989) 752;
S. Maedan and T. Suzuki, Prog.Theor.Phys. {\bf 80} (1988) 929;
S. Maedan $et$ $al$, Prog.Theor.Phys. {\bf 84} (1990) 130.
%17
\bibitem{Bere} V.L.~Beresinskii, {\it Sov.~Phys.~JETP} {\bf 32}
(1970) 493.
%18
\bibitem{KT} J.M.~Kosterlitz and D.J.~Thouless, {\it J.~Phys.},{\bf
C6} (1973) 1181.
%19
\bibitem{misha1} M.N. Chernodub and M.I. Polikarpov, unpublished.
%19'
\bibitem{Montovay} I.~Montvay and G.~M\"unster, ``Quantum Fields on a 
Latiice'' Cambridge University Press.
%20
\bibitem{teper98} M. Teper, hep-th/9812187.
%21
\bibitem{bali96} G.S.~Bali {\it et al.}, Phys. Rev. {\bf D54} (1996) 2863.
%22
\bibitem{stack} J.D. Stack, S.D. Nieman and R.J. Wensley, 
Phys. Rev. {\bf D50} (1994) 3399.
%23
\bibitem{JF} J.~Fingberg, U.~Heller and F.~Karsch,
{\it Nucl.~Phys.} {\bf B392} (1993) 493.
\end{references}
\end{document}